\documentclass[12pt]{article}

\usepackage{amsmath} 
\usepackage{amssymb}
\usepackage[T1]{fontenc}
\newcommand{\bea}{\begin{eqnarray}}
\newcommand{\eea}{\end{eqnarray}}
\newcommand{\bsa}{\begin{subeqnarray}}
\newcommand{\esa}{\end{subeqnarray}}

\newcommand{\Pe}{P\!e\>}

\newcommand{\be}{\begin{equation}}
\newcommand{\ee}{\end{equation}}

\newcommand{\bd}{\begin{description}}
\newcommand{\ed}{\end{description}}

\renewcommand{\Re}{{\rm Re}}

\newcommand{\sgn}{\text{\it {sgn}}}

\def\be{\begin{equation}}
\def\ee{\end{equation}}

\usepackage{graphicx}

\begin{document}

\title{On the symmetry properties of a random passive scalar with and without boundaries, and their connection between hot and cold states}
\date{\vspace{-5ex}}
\author{Roberto Camassa, Zeliha Kilic and Richard M. McLaughlin}

\maketitle
\begin{center} \normalsize{Department of Mathematics, University of North Carolina, Chapel Hill,}\\
\normalsize{Chapel Hill, North Carolina, 27599, USA}\\
\end{center}


\begin{abstract}
{We consider the evolution of a decaying passive scalar in the presence of 
a gaussian white noise fluctuating linear shear flow known as the Majda Model. We focus on deterministic 
initial data and establish the short, intermediate, and long time symmetry
properties of the evolving point wise probability measure (PDF) for the random
passive scalar.  We identify, for the cases of both point source and line source initial data, regions in the x-y plane outside of which the PDF skewness is sign definite for all time, while inside these regions we observe multiple sign changes corresponding to exchanges in symmetry between hot and cold leaning states using  exact representation formula for the PDF at the origin, and away from the origin, using numerical evaluation of the exact available Mehler kernel formulas for the scalars statistical moments.   A new, rapidly convergent Monte-Carlo method is developed, dubbed Direct Monte-Carlo (DMC), using the available random Green's functions which allows for the fast construction of the PDF for single point statistics, as well as multi-point statistics including spatially integrated quantities natural for full Monte-Carlo simulations of the underlying stochastic differential equations (FMC).  This new method demonstrates the full evolution of the PDF from short times, to its long time, limiting and collapsing universal distribution at arbitrary points in the plane.  Further, this method provides a strong benchmark for FMC and we document numbers of field realization criteria for the FMC to faithfully compute this complete dynamics.  Armed with this benchmark, we apply the FMC to a channel with a no-flux boundary condition enforced on parallel planes and observe a dramatically different long time state resulting from the existence of the wall.  In particular, the channel case collapsing invariant measure has {\em negative} skewness, with random states heavily leaning heavily towards the hot state, in stark contrast to free space, where the limiting skewness is positive, with its states leaning heavily towards the cold state.}

\end{abstract}

\section{Introduction}


The problem of computing the point-wise probability distribution function for a diffusing scalar advected by a random fluid flow is a challenging problem which has received considerable attention in the literature \cite{Majda,Pope,Kraichnan}. This problem emerged in the 1990s as a cartoon for fluid turbulence, as the passive scalar evolution possesses moment closure problems similar to those arising in fluid turbulence, and has been demonstrated to produce strongly non-Gaussian statistics in the presence of Gaussian velocity distributions \cite{Majda,McLMajda,BronskiMcL1,BronskiMcL2,CamassaLinMcL1,CamassaLinMcL2,Vanden} not unlike the observations of non-Gaussian, exponential like heavy temperature tails of the Chicago experiments of fluid turbulence from the late 1980s \cite{Libchaber}.  Generally, those studies were focused upon computing the infinite time, limiting invariant distributions, often utilizing random initial data to obtain the limiting dynamics efficiently.  

Here, we focus on the evolution at short, intermediate, and long time, with the ultimate goal being to assess the role boundaries play in affecting the statistical properties the scalar inherits from a randomly moving wall.  While still simpler than the full turbulence problem, our study is aimed at improving our understanding of how randomness evolves in variable coefficient partial differential equations, with particular attention payed to generating verifiable physical velocity profiles.  Our recent findings have stressed the importance of taking into account the influence of proper boundary conditions and the new, unexpected phenomena they lead to. In particular, motivated by our recent results in the laminar flow regime which documented the surprising role the tube geometry plays in the upstream/downstream symmetry properties of diffusing solutes transported by laminar flow \cite{Science,PRL}, we here explore the evolving statistical symmetry properties of a diffusing passive scalar advected by rapidly fluctuating, (Gaussian, white in time), linear shear flow with deterministic initial data.  While similar in spirit, our prior studies examined the first 4 spatial moments, $\int y^j T(x,y,z,t) dy$, for a passive scalar $T$, with $y$ the direction of the shear flow, $j=1,2,3,4$.  Here, we are interested in the statistical moments, $\langle T^j\rangle$, where the brackets refer to ensemble average over the statistics of the flow.  This analysis will develop a stringent benchmarking procedure which successfully documents the ability of a full Monte-Carlo simulation to capture the free space dynamics (for which much exact analysis is available).  With this benchmark complete, we present the results of Monte-Carlo simulations for the random passive scalar in the presence of a no-flux boundary condition enforced on a single plane in a half-space and document how the wall gives rise to a dramatically different limiting probability measure.  

The paper is organized as follows: In section one, we set up the problem, and briefly summarize the known results for the inherited scalar probability measure induced by this random flow.  In section 2, we present the problem setup and a discussion of the current literature on this problem.  In section 3, we present the random Green's function for the cases of line source and point source initial data.  We establish that the long time, universal invariant, limiting and collapsing probability measure is set by the behavior at the origin $x=y=0$.  For this point, we obtain exact formula for the PDF, and present its dynamic self-similar evolution properties.   In section 4, we develop the evolution properties of the statistical skewness in the $x-y$ plane. We first show that sufficiently far from the support set of the initial data, the skewness is positive for all time through a careful asymptotic analysis of the N dimensional Mehler kernel.  We then study the evolution of the PDF for line and point source cases using this method, and identify regions in the x-y plane where the skewness switches signs multiple times.  These regions are carefully determined using numerical integrations of the Mehler kernel, and provide strong validation of the DMC method.  In section 5, we develop the new Direct Monte-Carlo method which uses the temporal rescaling properties of Brownian motion, Brownian bridge and their L2 norms to quickly generate field realizations of the random passive scalar at any point in the $x-y$ plane through the explicit random Green's function.  In section 5, we turn to the comparison of DMC with full Monte-Carlo simulation of the underlying stochastic differential equations in free space.  We develop random representations of various integral quantities of the passive scalar, which are most natural for practical experimental applications involving optical concentration measurements and for Full Monte-Carlo simulations.  We establish the ability of the Full Monte-Carlo simulations to match the predictions of the DMC method for point source initial data, using one of these integral statistics from short, to the the long time, universal distribution.  Armed with this strong benchmark, in section 6, we present results of the Full Monte-Carlo Simulations to study the evolution of the probability distribution for this integrated passive scalar in free space and a channel with two parallel walls on which vanishing Neumann boundary conditions are enforced.  We will see a dramatic difference between the free space and channel domain case.  Lastly, in section 7, we discuss the results, and plans for a future experimental campaign.

\section{Formulation, background, and new processes in turbulent transport}
We consider the Majda Model for a passive scalar governed by a non-dimensional, random advection diffusion equation with deterministic initial condition $T_{0}\left(x,y,x_{0}\right)$:
\begin{eqnarray} 
&\frac{\partial T}{\partial t}+Pe\gamma(t)x\frac{\partial T}{\partial y}=\Delta T\\
&T(0,x,y,x_{0})=T_{0}(x,y,x_{0})
\label{passivescalar}
\end{eqnarray}
Here $\gamma(t)$ is Gaussian white noise  with the correlation function $R(t,s)$ defined as $\langle\gamma(t)\gamma(s)\rangle_{\gamma}=R(t,s)=\delta(t-s)$ and $Pe$ is standing for the non-dimensional {\em Peclet} number and we consider $T_{0}(x,y,x_{0})$ as 
\begin{equation*}
T_{0}^{L}{(x,y,x_{0})}=\delta\left(y\right) \quad \text{and}\quad T_{0}^{P}\left(x,y, x_{0}\right)=\delta\left(x-x_{0}\right)\delta\left(y\right)
\end{equation*}
such that the superscripts $L,P$ stand for line initial data and point source, respectively.  

The time dependent random velocity field can originate from either a time varying pressure field, or by randomly moving portions of the boundary.  To obtain the shear flow generically, requires solving a time varying heat equation with boundary conditions matching the wall motion (no-slip).  To illustrate this and bring forth the salient features of the boundary influence on the flow and tracer evolution, consider the case of the flow between two parallel plates which gives rise to a time varying shear layer, $u$:
\begin{eqnarray*}
&&u_{t}=\nu u_{xx},\qquad u(0,t)=0,\qquad u(L,t)=\gamma(t)\\
&&u(x,0)= 0,\qquad \langle\gamma(t)\gamma(s)\rangle=g(|t-s|).
\end{eqnarray*}
Here, $\gamma(t)$ could be any random process, such as white in time or Ornstein-Uhlenbeck process, with general time correlation function $g$, and the brackets henceforth refer to ensemble average.    
We can consider the following Fourier cosine series expansion for the random process $\gamma(t)$ (here on the time interval $[0,1]$,
$\gamma(t)=\sum_{j=0}^{\infty}{a_{j}\cos(\pi j t)}$, where the $a_j$ are Gaussian random numbers which, for the white in time case, have variance $\langle a_0^2\rangle=1$, and $\langle a^2_j\rangle=2$ for $j\neq 0$.  
The solution obtained by Laplace transform, taking the transform of the boundary motion to be $\hat f(s)$ is:
\begin{eqnarray*}
u(x,t)&=&\frac{1}{2\pi \text{i}}\int_{\mathit{C-\textit{i}\infty}}^{C+\textit{i}\infty}ds e^{st}\hat{f}(s)\frac{\sinh\left(\sqrt{\frac{s}{\nu}}x\right)}{\sinh\left(\sqrt{\frac{s}{\nu}}L\right)}\quad \text{such that }\quad C>\Re(\text{singularities})\\
\end{eqnarray*}
We can observe that the flow originally considered by Majda and collaborators \cite{Majda,McLMajda} is obtained in the limit of large viscosity, in this bounded setup,
\begin{eqnarray*}
\lim_{\nu \rightarrow \infty} u(x,t)= \gamma(t)\frac{x}{L},
\end{eqnarray*}
but in general we have the proper flow solution for arbitrary viscosities and wall motion which can be used in Monte-Carlo simulations when analytical approach to the scalar problem proves unwieldy.   Note that more general domains can be handled in a similar fashion provided the associated elliptic problem possesses an explicit eigenbasis; also note that much more general fluctuations may be similarly considered.  For the rest of this paper, we restrict our attention to the case with $g(t)=\delta(t)$, the so-called ``white noise limit".

While many passive scalars experience extremely complicated Peclet number dependence, for our case involving a linear shear flow, in the absence of boundaries, with white-in-time fluctuations, the Peclet dependence may be completely rescaled from the problem via:  $(x',y')= (Pe x,Pe y)$, and $(t'=Pe^2 t)$, and as such, unless otherwise noted, we will set $Pe=1$.  Of course, with the addition of physical boundaries, the Peclet number will re-enter the problem.  

We give a short review of the history of this problem next.

Majda \cite{Majda} established by directly computing the ensemble average of the Feynman-Kacs functional representation that the general $N$ point statistical correlator, $\psi_N=\left<\prod_{j=1}^N T(x_j,y_j,t)\right >$, satisfies a close partial differential equation in $2N$ dimensions (here brackets refer to the ensemble average of the random velocity field), and further established that the streamwise Fourier transform of the $N$ point statistical scalar correlator, $\Psi = \langle \prod_{j=1}^N \hat T(x_j,k_j,t)\rangle$ satisfies an $N$-body parabolic quantum mechanics problem \cite{Majda, McLMajda}:
\begin{eqnarray*}
\frac{\partial \Psi}{\partial t} &=& \Delta_N \Psi -V(\boldsymbol{x},\boldsymbol{k}) \Psi\\
V(\boldsymbol{x},\boldsymbol{k})  &=& -4 \pi^2 \kappa |\boldsymbol{k}|^2 - 2 \pi^2 (\boldsymbol{x} \cdot \boldsymbol{k})^2\\
\Psi(\boldsymbol{x},\boldsymbol{k},0)&=& \delta (\boldsymbol{x}-\boldsymbol{x_0})
\end{eqnarray*}
where $\Delta_N$ is the Laplacian operator in $N$ dimensions (for the above flow), $\Delta_N=\sum_{j=1}^N \partial^2_{x_j}$, $\boldsymbol{k}=(k_1,k_2,\cdots,k_N)$, and $\boldsymbol{x}=(x_1,x_2,\cdots,x_N)$.  
For initial data depending only on the streamwise coordinate, Majda derived exact solutions in any dimension for this problem \cite{Majda}, while for more general multi-dimensional initial data, McLaughlin and Majda \cite{McLMajda,McLthesis} utilized rotations of coordinates in $R^N$ to derive explicit formulae for the wavefunction.  Majda established that for Gaussian random initial data depending on the streamwise coordinate, that the inherited scalar distribution would develop broader than Gaussian flatness factors for any positive time.  It was then shown that this property holds for more general initial data \cite{McLMajda}.  McLaughlin and Majda established for deterministic initial data, non-zero, positive skewness is observed in the limiting probability measure.  This will correspond to the {\em cold state} leaning invariant measure discussed below in free-space.

Bronski and McLaughlin computed the large $N$ asymptotics for this problem \cite{BronskiMcL2} and computed rigorous estimates for the decay rate of the tail of the probability distribution.  They also demonstrated how the tail depends upon the initial data power spectrum, specifically showing how as the energy in the initial data is moved to smaller scales, the PDF develops a broader and broader tail, consistent with the moments analysis Majda performed.  Later, Vanden-Eijnden extended these results, primarily in the long time limit, to a broader class of non-white in time Gaussian random shear flows with finite or even infinite correlation times \cite{Vanden}.  

These results were all derived in free-space, in the absence of physical boundaries.  Much less is known for random partial differential equations in the presence of physical boundaries.  Bronski and McLaughlin \cite{BronskiMcL1} utilized second order perturbation theory for the ground state of periodic Schrodinger equations to analyze the inherited probability measure for a passive scalar advected by periodic shear flows with multiplicative white noise, assuming the scalar field satisfied periodic boundary conditions.  In that work, they established for random initial data depending upon the stream-wise variable that a Central Limit theorem dominated at long times, with non-uniform relaxation of the moments to their Gaussian numbers.  For deterministic initial data, with physical (vanishing Neumann) boundary conditions, essentially nothing is known, particularly on finite timescales, and understanding differences in this setting is the ultimate goal of this paper.  The maximum principle for the passive scalar insures that for any positive time, with deterministic initial data, the ensuing probability measures will be compactly supported, with a sharp estimate derived below for this support set.  Further, as we are interested ultimately in experimental observations, we will focus on positive initial data, and we will be most interested in assessing the skewness which will indicate if the tracer fluctuations are biased towards a hot or cold state.  

Towards this goal, we will first develop new random Green's function formulae for the two types of initial data of line and point source outlined above.  This will be done using the method of characteristics and careful back-Fourier transform analysis for the case of the point source.  We will observe the emergence of the following stochastic processes in this analysis, which we present now.

Let $B(s)$ be the standard Brownian motion and $\eta=\int_{0}^{1}{B^{2}(s)ds}$ is the $L^{2}$ norm of Brownian motion.  The PDF of the $L^{2}$ norm of Brownian motion as well as the
single point statistics of the scalar field corresponding to the random initial data case was studied in \cite{CamassaLinMcL2}. 
Due to the rescaling property of the Wiener process, we can write the $L^{2}$norm of Brownian motion \cite{CamassaLinMcL2,Vanden} and the Brownian Bridge, $\mu,$ as follows:
\begin{eqnarray}
\xi(t)&=&\int_{0}^{t}{duB^{2}\left(u\right)}\equiv t^{2}\int_{0}^{1}{ds B^{2}(s) }=t^{2}\eta\quad \text{in law}\\
\eta&=&\int_{0}^{1}du B^{2}(u)\\
\mu&=&\int_{0}^{1}{du \left[B(u)-\int_{0}^{1}dsB(s)\right]^{2}}
\label{law}
\end{eqnarray}
It was established by Camassa, et. al \cite{CamassaLinMcL2} how the $L^{2}$ norm of Brownian motion affects the random behavior of the scalar field with random initial data, and in particular, they observed an interesting dynamic phenomena in which the evolving PDF exhibited a non-monotonic approach to the invariant long time measure for certain classes of initial data.  Next, we extend those results to the more complicated case involving deterministic initial data.  Surprisingly, we will see that the L2 norm of the Brownian Bridge emerges in this model for the first time for the case of deterministic point source initial data.

\section{Random Green's functions and universal invariant measures at long time }

In this section, we calculate the explicit random Green's function for line and point source initial data. We then develop exact integral expressions for the long time invariant measures, and calculate asymptotic expansions for this measure for its small and large fluctuation values.  We demonstrate how the PDF at finite time for the temperature at the origin evolves to this limiting state on short, intermediate, and long times.  

\subsection{Line source initial data}
Consider the following pde 
\begin{eqnarray} 
\frac{\partial T^{L}}{\partial t}+\gamma(t)x\frac{\partial T^{L}}{\partial y}&=&\Delta T^{L}\\
T^{L}(0,x,y,x_{0})&=&\delta (y)
\label{PDEline}
\end{eqnarray}
It is an elementary application of the method characteristics to calculate the random Green's function associated with line source initial data:
\begin{eqnarray}
T(t,x,y)&=&\frac{\exp{\big(\frac{-(y-xB(t))^{2}}{4t(1+t\int_{0}^{1}{ds B^{2}(s)})}\big)}}{\sqrt{4\pi t(1+t\int_{0}^{1}{ds B^{2}(s)})}}\;\; .
\label{RGLine}
\end{eqnarray}
In this formula, we have utilized the well known time rescaling properties for the L2 norm of Brownian motion given in equation (\ref{law})
This has a particularly simple form when evaluated at the origin $x=y=0$:  The random field at this point is a nonlinear mapping of the $L2$ norm of Brownian motion.  Consequently, elementary probability theory can be used to construct the PDF for scalar at this point, using the known PDF for the L2 norm of Brownian motion \cite{CamassaLinMcL2}: 
\begin{eqnarray}
P_{\eta}(\eta)&=&\frac{1}{2\pi \textit{i}}\int_{\Gamma-\textit{i}\infty}^{\Gamma+\textit{i}\infty}{ds \frac{\exp(s\eta)}{\big[\prod_{k=1}^{\infty}\big(1+\frac{2s}{(k-\frac{1}{2})^{2}\pi^{2}}\big)\big]}}\\
&=&\frac{1}{\pi}\sum_{k=0}^{\infty}{(-1)^{k} e^{\left({{-\left(2k+\frac{1}{2}\right)^{2}\frac{\pi^{2}}{2}\eta}}\right)}}\int_{0}^{(2k+1)\pi^{2}}dr\frac{\exp(-r\eta)}{\sqrt{-\cos\left(\sqrt{\big(2(r+(2k+\frac{1}{2})^{2}\frac{\pi^{2}}{2})\big)}\right)}}
\label{loopline}
\end{eqnarray}
The PDF for the random variable,  $z=T^{L}(t,0,0,0)$, is immediately available from this expilcit PDF formula through direct change of variables:
\begin{eqnarray}
P(z)=\frac{1}{2\pi t^{2}z^{3}}P_{\eta}\left(\frac{1}{t }\left(\frac{1}{4\pi tz^{2}}-1\right)\right)
\label{mappedmeasure}
\end{eqnarray}
Since $z$ is a positive random variable, and since $\eta$ is a positive random variable, we have immediately from equation (\ref{mappedmeasure}) an estimate for the support set of the $z$:  $0\le z\le 1/\sqrt{4 \pi t}$.  This is also immediately evident from inspection of the random Green's function solution in  (\ref{RGLine}) generally for $T$ even for points away from the origin since the argument of the exponential is positive.  But this support estimate is not sharp:  The measure collapses faster, and the invariant, long time measure can be immediately deduced from rescaling $z$ by:
\begin{eqnarray}
\tilde z&=&t \sqrt{4\pi}  z = \frac{t \sqrt{4\pi}  }{\sqrt{4\pi t ( 1 + t\eta)}}\\
&\sim& \frac{1}{\sqrt{\eta}} \;\;, \; t\to \infty
\end{eqnarray}
This invariant measure can be immediately constructed again using the PDF for the L2 norm of Brownian motion through change of variables.  To this end: 
\begin{eqnarray}
P(\tilde z) &=& \frac{2}{\tilde z^3} P_{\eta} (1/\tilde z^2)
\end{eqnarray}

\begin{figure}[h]
\centering
\includegraphics[width=0.45\textwidth]{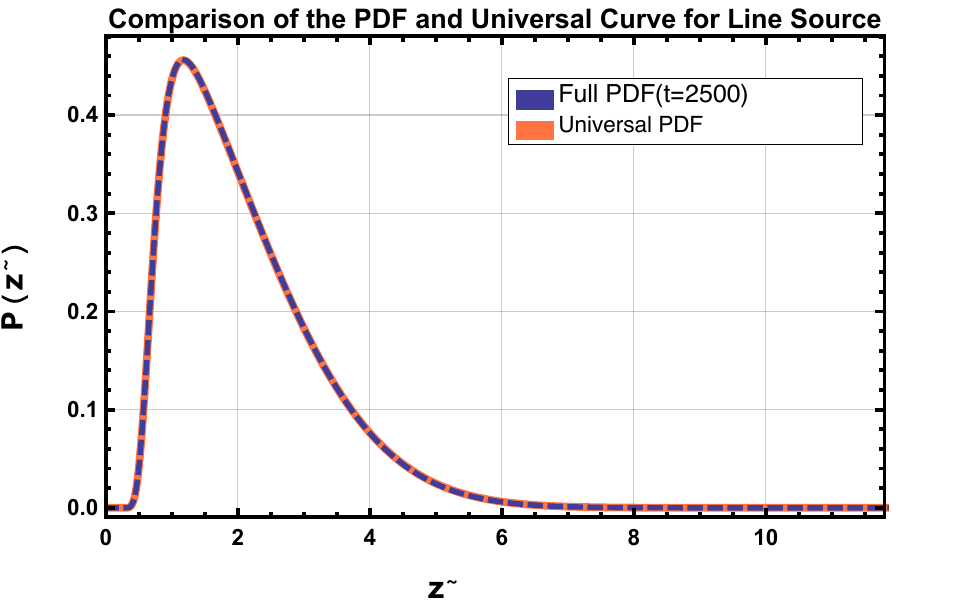}
\includegraphics[width=0.45\textwidth]{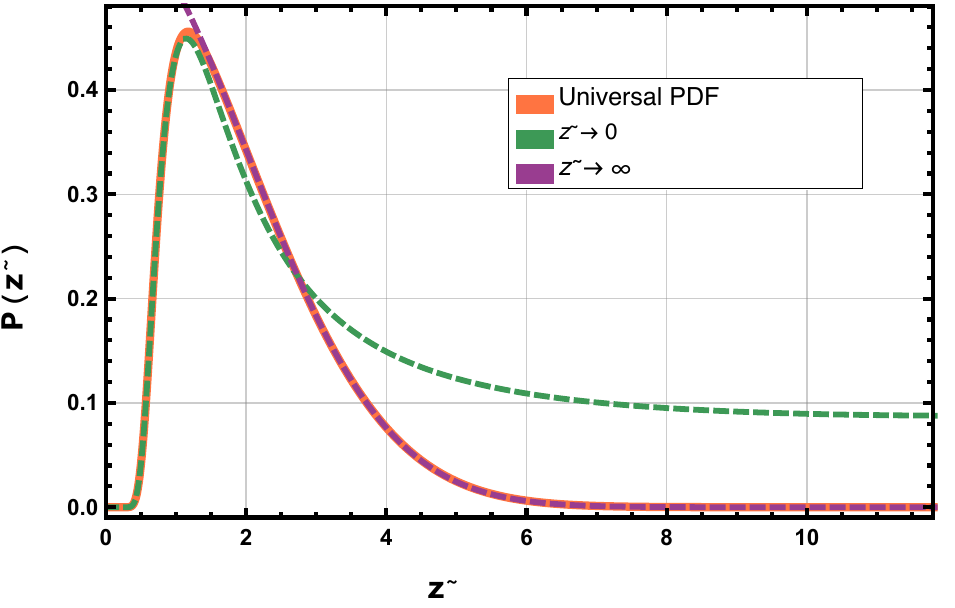}
\caption{{\footnotesize Limiting invariant measure for $\tilde z$ for line source initial data: Left panel,  comparison between full PDF retaining 21 terms in (\ref{loopline}) at time $t=2500$ (yellow) and limiting invariant measure (grey) using 31 terms. Right panel: comparison between invariant measure and small $\tilde z$ asymptotics (long dash), and large $\tilde z$ asymptotics (short dash) both evaluated at time $t=2500$. }}
\label{univline}
\end{figure}

Shown in figure \ref{univline} is this limiting probability measure drawn through numerical evaluation of the loop integrals in equation (\ref{loopline}) retaining $31$ terms.   We remark that while the series is alternating, the number of terms necessary for accurate evaluation basically requires $k_{max} \gg \tilde z$.  Observe that the maximum of this distributions at approximately, $\tilde z = 1.2$.  Also shown is the PDF evaluated at time $t=2500$, which  has converged to the long time limit.  In the original coordinates, this maximum will collapse to the origin at rate $1/t$, and the peak height scales linearly in time, establishing that the measure collapses to a delta limiting sequence supported at the origin in long time, with cold leaning (positive skewness) states.   

\begin{figure}[h]
\centering
\includegraphics[width=0.7\textwidth]{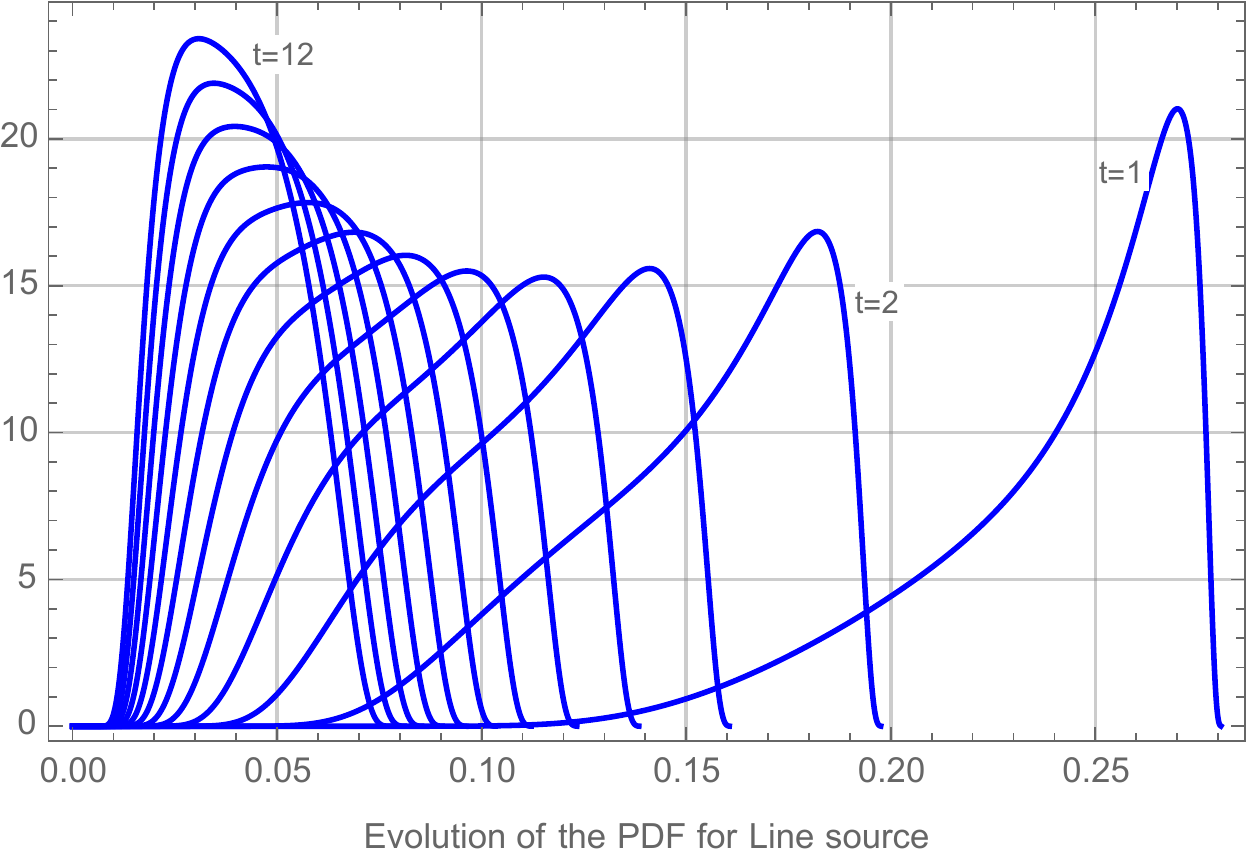}
\caption{{\footnotesize Evolution of the PDF for the line source for the single point statistics of the scalar field $T$ at $(x,y)=(0,0)$ by using the exact PDF formula in equation (\ref{loopline}).}}
\label{lineevolution}
\end{figure}

The asymptotic properties of the invariant measure for $\tilde z$ are derived in two different ways.  For the case of small values, it is an immediate application of Watson's lemma applied to the first term in series of mapped loop integral defining the complete PDF:
\begin{equation}
P(\tilde z)\sim\frac{\sqrt{2} \sqrt{t} e^{-\frac{1}{8} \pi ^2 \left(\frac{1}{{\tilde z}^2}-\frac{1}{t}\right)} \left(\frac{t {\tilde z}^2}{2 \pi ^2 \left(t-{\tilde z}^2\right)}+1\right)}{{\tilde z}^2 \sqrt{t-{\tilde z}^2}} \; ,\quad \tilde z \rightarrow 0 \, .
\label{asymp1}
\end{equation}
We remark that this formula is valid for all time, and evaluating it at long times provides a strong agreement with the invariant measure shown in figure \ref{univline}.  

Alternatively, the behavior for large values of $\tilde z$ is more complicated involving a super-exponential ansatz following prior work of two of the authors \cite{CamassaLinMcL2}:
\begin{equation} 
P(\tilde z)\sim\frac{t^{3/2} e^{-\frac{t\tilde z^{2}}{8
   \left(t-\tilde z^{2}\right)}}}{\sqrt{\pi }
   \sqrt{\left(t-{\tilde z}^2\right)^3}} \; , \quad \tilde z\rightarrow \infty \;\;.
\end{equation}
We remark that this formula is valid only under the assumption of large time and large $\tilde z$, with the restriction: $1\ll \tilde z \ll \sqrt{t}$.  
These formulae are plotted in figure \ref{lineevolution} and even overlap capturing the peak.  We remark that sending time to infinity in equation (\ref{asymp1}) does not capture the peak of the invariant measure as accurately, though is still asymptotically valid for small $\tilde z$.  
We will see below that the analogous formulae for the point source case do not capture the peak at leading order, even retaining finite time, for the small $z$ expansion.

The loop integral formula in equation (\ref{loopline}) combined with the mapping in (\ref{mappedmeasure}) allow for a rapid evaluation of the scalar at the origin using numerical evaluation of these loop integrals.  In figure \ref{lineevolution} we show the evolution of this measure on short, intermediate, and long timescales.  
Observe that the skewness of these measures starts negative, but transitions through a symmetric state at approximate time $t=8.157$, before ultimately collapsing into a delta limiting sequence with positive skewness on long times.  This ultimately collapses to the invariant measure in rescaled coordinates for $ \tilde z$.  

Lastly, we stress that inspection of the random Green's function given in (\ref{RGLine}) shows that the for any point in the plane, that the long time pointwise PDF will always converge eventually to this universal state:  a delta limiting sequence with positive skewness corresponding to a cold-leaning biased state. We stress that this property also holds for the case of point source initial data, with the only change being in the timescale to access the final state.  Below in section 4, we will examine finite time behavior for general points in the plane to see some interesting dynamics which occur before this final collapsing state is reached.

\subsection{Point source initial data}

We next consider the behavior of the random passive scalar for the case of a point source initial data.  In this case:
\begin{eqnarray} 
\frac{\partial T^{P}}{\partial t}+\gamma(t)x\frac{\partial T^{P}}{\partial y}&=&\Delta T^{P}\\
T^{P}(0,x,y,x_{0})&=&\delta(y)\delta(x-x_0)
\label{PDEPoint}
\end{eqnarray}

Obtaining the explicit formula for random Greens function for this initial data is a bit more involved than the line source case, requiring a smart choice for rotating coordinates to correctly evaluate the inverse Fourier transform.

After applying Fourier transform in both directions to equation (\ref{PDEPoint}) we solve the resulting problem by method of characteristics. This yields \begin{eqnarray*}
{T^{P}}(t,x,y,x_{0})&=&\int_{\mathbb{R}^{2}}{dkd\eta}{e^{2\pi \textit{i}(kx+\eta y)}e^{-2\pi \textit{i}x_{0}(k+\eta\textit{Pe}B(t))}}e^{-4\pi^{2}\int_{0}^{t}{du}\left(\eta^{2}+\left(k+\eta\textit{Pe}\left(B(t)-B(u)\right)^{2}\right)\right)}
\end{eqnarray*}
After we make the following change of variables:
\begin{eqnarray}
k&=&r-B(t)q\textit{Pe}\\
\eta&=&q\\
\end{eqnarray}
At this step we have to integrate in $r$ first.  Otherwise, the resulting second integral in $r$ defines a random greens function with a non-conditionally positive variance.
\begin{eqnarray}
{T^{P}}(t,x,y,x_{0})=\int_{\mathbb{R}^{2}}{drdq}{e^{2\pi \textit{i}(r-B(t)q\textit{Pe})}e^{2\pi \textit{i}q y}e^{-2\pi i x_{0}r}}{e^{-4\pi^{2}q^{2}t}e^{-4\pi^{2}\int_{0}^{t}du\left(r-qB(u)\textit{Pe}\right)^{2}}}\\
=\int_{\mathbb{R}}{dq {e^{2\pi \textit{i} q\left(y- \textit{Pe}xB(t)\right)}}e^{-4\pi^{2}q^{2}\left(t+{\textit{Pe}}^{2}\int_{0}^{t}duB^{2}(u)\right)}}{\frac{e^{\frac{\left(2\pi\textit{i}\left(x-x_{0}\right)+8\pi^{2}q\textit{Pe}\int_{0}^{t}duB(u)\right)^{2}}{16\pi^{2}t}}}{\sqrt{4\pi t}}}\\
=\frac{e^{-\frac{(x-x_{0})^{2}}{4t}}}{\sqrt{4\pi t}}\frac{e^{\frac{-\left(y- \textit{Pe}xB(t)+(x-x_{0})\textit{Pe}\frac{\int_{0}^{t}duB(u)}{t}\right)^{2}}{4t\left(1+\left(\frac{{\textit{Pe}}}{t}\right)^{2}{\left(t\int_{0}^{t}duB^{2}(u)-{\left(\int_{0}^{t}duB(u)\right)^{2}}\right)}\right)}}}{\sqrt{4\pi t\left(1+\left(\frac{{\textit{Pe}}}{t}\right)^{2}{\left(t\int_{0}^{t}duB^{2}(u)-{\left(\int_{0}^{t}duB(u)\right)^{2}}\right)}\right)}}\\
\equiv\frac{e^{-\frac{(x-x_{0})^{2}}{4t}}}{\sqrt{4\pi t}}\frac{e^{-\frac{\big(y-B(t)\textit{Pe}x+\textit{Pe}\sqrt{t}(\int_{0}^{1}{duB(u)})(x-x_{0})\big)^{2}}{4t(1+t\textit{Pe}^{2}\int_{0}^{1}{du[B(u)-\int_{0}^{1}{dsB(s)}]^{2}})}}}{\sqrt{4\pi t(1+t\textit{Pe}^{2}\int_{0}^{1}{du[B(u)-\int_{0}^{1}{dsB(s)}]^{2}})}}
\label{RGPoint}
\end{eqnarray}
where the last equality (in law) takes advantage of the rescaling properties of Brownian motion.  We emphasize, however, that this calculation is valid for any stochastic process $B(t)$ except for the last equality, and it will be useful to explore different processes in the future using this approach

We now recognize the central stochastic processes for this setup:
\begin{eqnarray}
\mu&=&\int_{0}^{1}{du [B(u)-\int_{0}^{1}dsB(s)]^{2}}
\label{bridge}\\
\eta&=& \int_{0}^{1}{du B^{2}(u)}\\
\beta&=& \int_{0}^{1}{du B(u)}
\end{eqnarray}
Observe the emergence of the L2 norm of the Brownian Bridge in equation (\ref{bridge}), which to our knowledge has not been observed previously for the Majda Model.  

The PDF for the L2 norm of the Brownian bridge was studied in \cite{Anderson,Smirnov}, where a rapidly convergent series representation was derived:
\begin{eqnarray}
P_{\mu}(\mu)=&\mu^{-\frac{5}{4}}{\sum_{n=0}^{\infty}}\frac{(-1)^{n}}{n!}\exp\left(-\frac{(n+\frac{1}{4})^{2}}{\mu}\right)\frac{1}{\Gamma(\frac{1}{2}-n)}D_{\frac{3}{2}}\left(\frac{2n+\frac{1}{2}}{\sqrt{\mu}}\right)
\label{looppoint}
\end{eqnarray}
where $D_{3/2}$ is the Parabolic Cylinder function of order $3/2$.  We may similarly construct the exact PDF for the random scalar field at the origin for the special case of $x_0=0$ using elementary mapping:
\begin{eqnarray}
P(z)=\frac{1}{8\pi^{2}t^{3}z^{3}}P_{\mu}\left(\frac{1}{t}\left(\frac{1}{16\pi^{2}t^{2}z^{2}}-1\right)\right)
\label{mappedmeasurepoint}
\end{eqnarray}

Like in the prior case for the line source, renormalization of the scalar field at the origin with $x_0=0$ yields the limiting invariant measure:

\begin{eqnarray}
\bar z&=&t^{3/2} \sqrt{4\pi}  T^P(x,y,t,0) = \frac{t \sqrt{4\pi}  }{\sqrt{4\pi t ( 1 + t\mu)}}\\
&\sim& \frac{1}{\sqrt{\eta}} \;\;,\; t\to \infty
\end{eqnarray}
This invariant measure can be immediately constructed again using the PDF for the L2 norm of the Brownian Bridge through change of variables.  To this end: 
\begin{eqnarray}
P(\bar z) &=& \frac{2}{\bar z^3} P_{\mu} (1/\bar z^2)
\end{eqnarray}

\begin{figure}[h]
\centering
\includegraphics[width=0.45\textwidth]{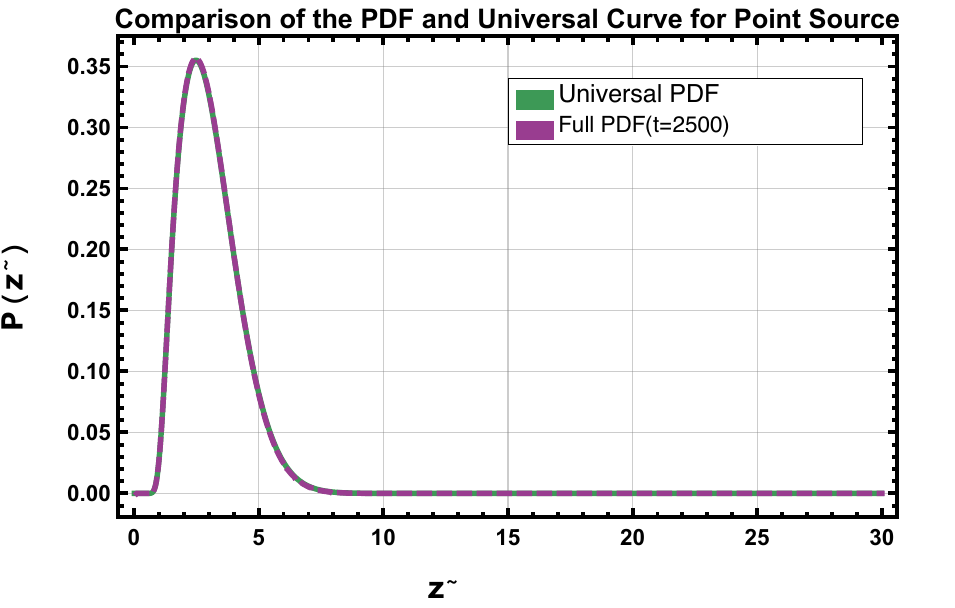}
\includegraphics[width=0.43\textwidth]{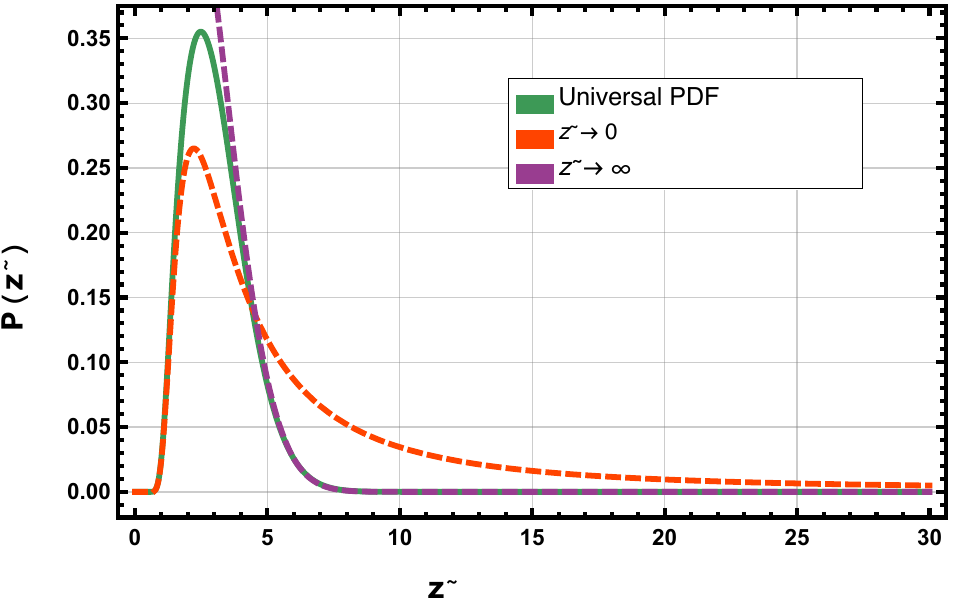}
\caption{{\footnotesize Limiting invariant measure for $\bar z$ for line source initial data: Left panel,  comparison between full PDF retaining 21 terms in (\ref{looppoint}) at time $t=2500$ (yellow) and limiting invariant measure (grey) using 31 terms. Right panel: comparison between invariant measure and small $\bar z$ asymptotics (long dash), and large $\bar z$ asymptotics (short dash) both evaluated at time $t=2500$. }}
\label{univpoint}
\end{figure}

Shown in figure \ref{univpoint} is this limiting probability measure drawn through numerical evaluation of the loop integrals in equation (\ref{looppoint}) retaining $30$ terms.  The same discussion above regarding the number of terms retained applies, with the estimate now being for this integral representation $k_{max} \gg 1/\bar z$.
Observe that the maximum of this distributions at approximately, $\bar z = 2$.  In the same manner as the case of the line source, in the original coordinates, this maximum will collapse to the origin at rate $1/t$, and the peak height scales linearly in time, establishing that the measure collapses to a delta limiting sequence supported at the origin in long time, with cold leaning (negative skewness) states.   

\begin{figure}[h]
\centering
\includegraphics[width=0.7\textwidth]{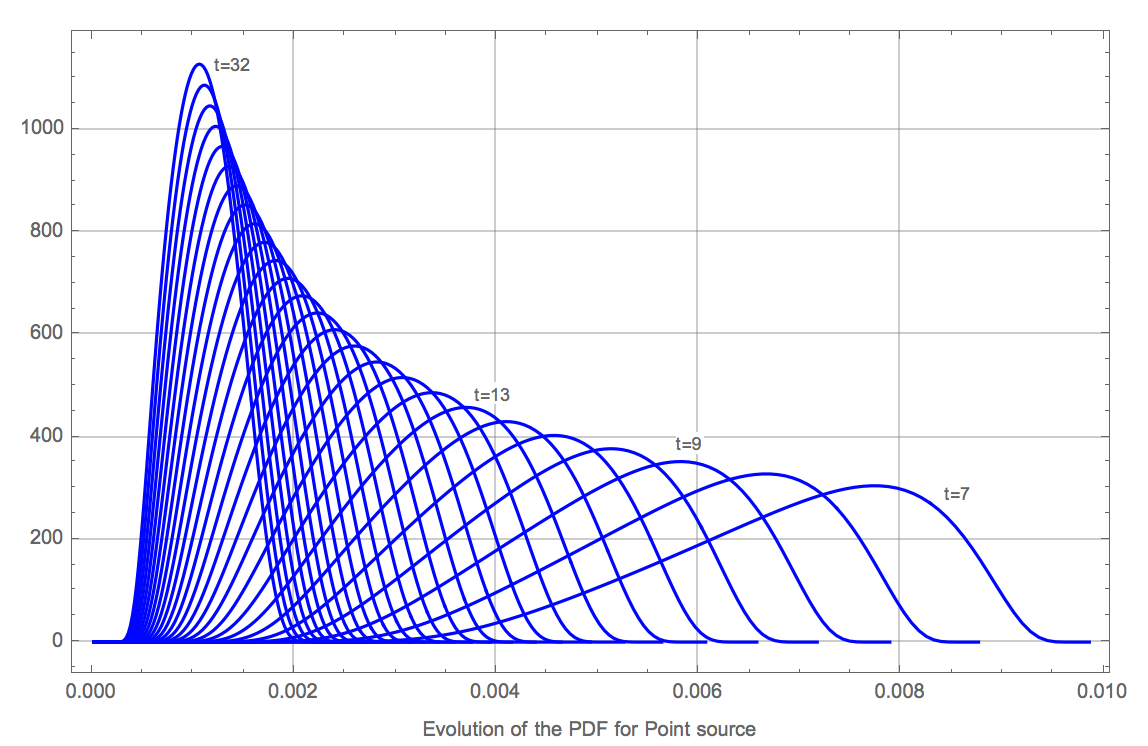}
\caption{{\footnotesize Evolution of the PDF for the point source for the single point statistics of the scalar field $T$ at $(x,y)=(0,0)$ by using the exact PDF formula in equation (\ref{loopline}).}}
\label{pointevolution}
\end{figure}

The asymptotic properties of the invariant measure for $\bar z$ are derived in two different ways.  For the case of small values, the asymptotics are derived using the well known large argument asymptotics for Parabolic Cylinder Functions, needing to only retain the first term in the series for small $\bar z$:
\begin{eqnarray}
P(\bar{z})\sim\frac{2 \sqrt{\pi t} e^{\frac{1}{2} \pi ^2
   \left(\frac{1}{t}-\frac{1}{\bar{z}^2}\right)}}{\bar{z}^2 \sqrt{{t-\bar{z}^2}}}\;\;, \bar z \to 0\;.
\end{eqnarray}
We remark that this asymptotic expansion is valid for all $t>0$.  

Similar to the line source, the behavior for large values of $\bar z$ requires the super-exponential ansatz of Camassa, McLaughlin, and Lin \cite{CamassaLinMcL2}:  
\begin{eqnarray}
P(\bar{z})\sim\frac{t^2 \bar{z} e^{-\frac{t \bar{z}^2}{8\left( t-\bar{z}^2\right)}}}{\sqrt{2
   \pi } \left(t-\bar{z}^2\right)^2}\;\;,  \bar z \to \infty \; .
\end{eqnarray}
These asymptotic expansions are superimposed over the limiting invariant PDF for $\bar z$ shown in figure \ref{univpoint} evaluated at time $t=2500$.  We observe that these leading order asymptotic formulaes do not do quite as good a job, particularly failing to capture the peak, even when evaluating at longer times.  Presumably higher order corrections will succeed in producing a larger overlap region between the small and large asymptotics of the PDF.  

The infinite series formula in equation (\ref{looppoint}) combined with the mapping in (\ref{mappedmeasurepoint}) allow for a rapid evaluation of the scalar at the origin using numerical evaluation.  In figure \ref{pointevolution} we show the evolution of this measure on short, intermediate, and long timescales.  
Observe that the skewness of these measures starts negative, but transitions through a symmetric state at approximate time $t=22.69$, before ultimately collapsing into a delta limiting sequence with positive skewness on long times.  This ultimately collapses to the invariant measure in rescaled coordinates for $ \tilde z$.  

Next we turn to examine the evolution of the statistical skewness for the line and point source cases in the x-y plane.  We will see that these long time invariant measures are reached non-monotonically for regions of the plane, with the skewness changing signs multiple times before landing on its long time, positive, cold leaning state. 

\section{Evolution of statistical skewness in free space}

Here we examine using numerical evaluation of the exact Mehler kernel representation for the statistical skewness in free space for the cases of line and point source initial data.  We recall that McLaughlin and Majda \cite{McLMajda} established that the long time limit of the skewness, for any point in the plane, would be a positive number for deterministic initial data in free space.  
Here we examine the evolution of the skewness to this long time limiting state, and find some complex dynamics in which the PDF changes its symmetry properties multiple times before reaching its long time limiting, cold leaning state.  

To begin with, we give some intuition regarding the short time dynamics of skewness in this problem:  Since we are considering deterministic initial data, at time $t=0$, we may consider that the deterministic initial data has a statistical interpretation as a Dirac mass in space supported on the initial curve:  $P(T(x,y,0)=\omega) = \delta(\omega-T_0(x,y))$.  Immediately for $t>0$, the heat operator smooth's this distribution, and the random advection injects randomness into the solution. If the evaluation point $(x,y)$ is far from the line or point source, it should be expected that the probability of finding a value different than zero should be extraordinarily small on short timescales, with some sort of distribution localized very near the origin.  It is less clear how this distribution should be shaped, and the analysis of the skewness shows us that sufficiently far away from the support set of the initial data, this distribution will always have positive skewness for {\em all} times:

\subsection{Large $x$, large $y$ asymptotics for $T^{L}_{0}(x,y,x_{0})=\delta(y)$ initial data}
Here we present the key features of an asymptotic analysis for the skewness along the line $x=y$ for large values of y. This result, documented in the Appendix, establishes that in this limit, the skewness is positive for all positive time.  The proof of this fact relies on splitting the Mehler kernel in a very careful manner:  The general N'th moment for this problem is
\begin{eqnarray}
\langle T^N(x,y,t)\rangle &=& \int_{R^N} d\mathbf{k} \frac{e^{\Phi(x,y,\mathbf{k},t)}}{\cosh{\left(2\sqrt{2} \pi |\mathbf{k}| t\right)}}\\
\Phi&=& 2\pi i (\sum_j k_j) y -4\pi^2 |\mathbf{k}|^2 t - \frac{\pi x^2 (\sum_j k_j)^2}{\sqrt{2} |\mathbf{k}|}  \tanh{\left( 2\sqrt{2} \pi | \mathbf{k} | t\right)}
\end{eqnarray}
One may always find an invertible linear transformation with $u_1 = (\sum_j k_j)$, with $|\mathbf{k}|^2 = \sum a_j u_j^2$ with $a_j>0$.  In these new coordinates, the integral becomes:
\begin{eqnarray}
\langle T^N(x,y,t)\rangle &=& |J|\int_{R^N} d\mathbf{u} \frac{e^{\Phi(x,y,\mathbf{u},t)}}{\cosh{\left(2\sqrt{2} \pi \sqrt{\sum a_j u_j^2} t\right)}}\\
\Phi&=& 2\pi i u_1 y -4\pi^2 \sum a_j u_j^2 t - \frac{\pi x^2 u_1^2}{\sqrt{2} \sqrt{\sum a_j u_j^2}}  \tanh{\left( 2\sqrt{2} \pi \sqrt{\sum a_j u_j^2} t\right)}
\end{eqnarray}
where $J$ is the Jacobian of the transformation.
Next, rescale the $u_1$ integral by $w=y u_1$, and let $x=y$.  For large values of $y$, this results in the following asymptotic expression the N'th Moment:
\begin{eqnarray}
&&\langle T^N(x,y,t)\rangle \sim\frac{|J|}{y} \int_{R^N} d\mathbf{u}\frac{e^{\bar \Phi(w,\mathbf{u},t)}}{\cosh{\left(2\sqrt{2} \pi \sqrt{\sum_{j=2} a_j u_j^2} t\right)}}\\
&&\bar \Phi= 2\pi i w -4\pi^2 \sum_{j=2} a_j u_j^2 t - \frac{\pi w^2}{\sqrt{2} \sqrt{\sum_{j=2} a_j u_j^2}}  \tanh{\left( 2\sqrt{2} \pi \sqrt{\sum{j=2} a_j u_j^2} t\right)}
\end{eqnarray}
Now the integral in $w$ is an explicit Gaussian integral which may be immediately evaluated, producing a convergent $N-1$ dimensional integral depending only upon time.  The key observation regarding the centered third moment follows immediately in that since each moment will decay like $1/y$ for large $y$:
\begin{eqnarray}
\langle (T- \langle T\rangle )^3\rangle &=& \langle T^3\rangle - 3 \langle T^2\rangle \langle T\rangle + 2 \langle T \rangle^3\\
&\sim&  \langle T^3\rangle\\
&\sim&\frac{1}{y} F(t) \;, \; y \to \infty
\end{eqnarray} 
for an explicitly positive function $F(t)$ given in terms of an $N-1$ dimension integral presenteds in the appendix resulting from the explicit evaluation of the one dimension Gaussian integral.  This result confirms the intuition that sufficiently far from the support of the initial line source, the sign of the statistical skewness is always positive for all time.  Similar arguments may be developed for the point source, but are more complicated.

\subsection{Evolution for statistical skewness for line and point source initial data}

The evolution of the statistical skewnesss in the finite plane is quite rich.  The skewness is the normalized, centered third moment:  
$S(x,y,t) =\frac{\langle (T-\langle T\rangle)^3\rangle}  {\langle (T-\langle T \rangle)^2\rangle^{3/2} }  $.  
The Mehler kernel exact integral representation for this quantity is given below in the Appendix.  

We will show that the skewness experiences multiple sign changes within a region in the $x,y$ plane.  First some physical intuition can help to understand the origin of this behavior.  As discussed above, points far from the support set of the initial data will have values at short times which are likely to be very small:  the PDF for such points will appear as some sort of measure tightly localized near the zero value.  Once the action of random advection and molecular diffusion have sufficient time to act, the tightly localized initial data will propagate to the observation point, and the support of the PDF will move away from zero to some finite value.  Then once the heat wave passes, the measure will again collapse into the origin. How the measure is shaped, leaning towards the hot state, or towards the cold state is not so easy to predict, and requires the calculation of the PDF, or statistics like the skewness.  

\begin{figure}[h]
\centering
\includegraphics[width=1.\textwidth]{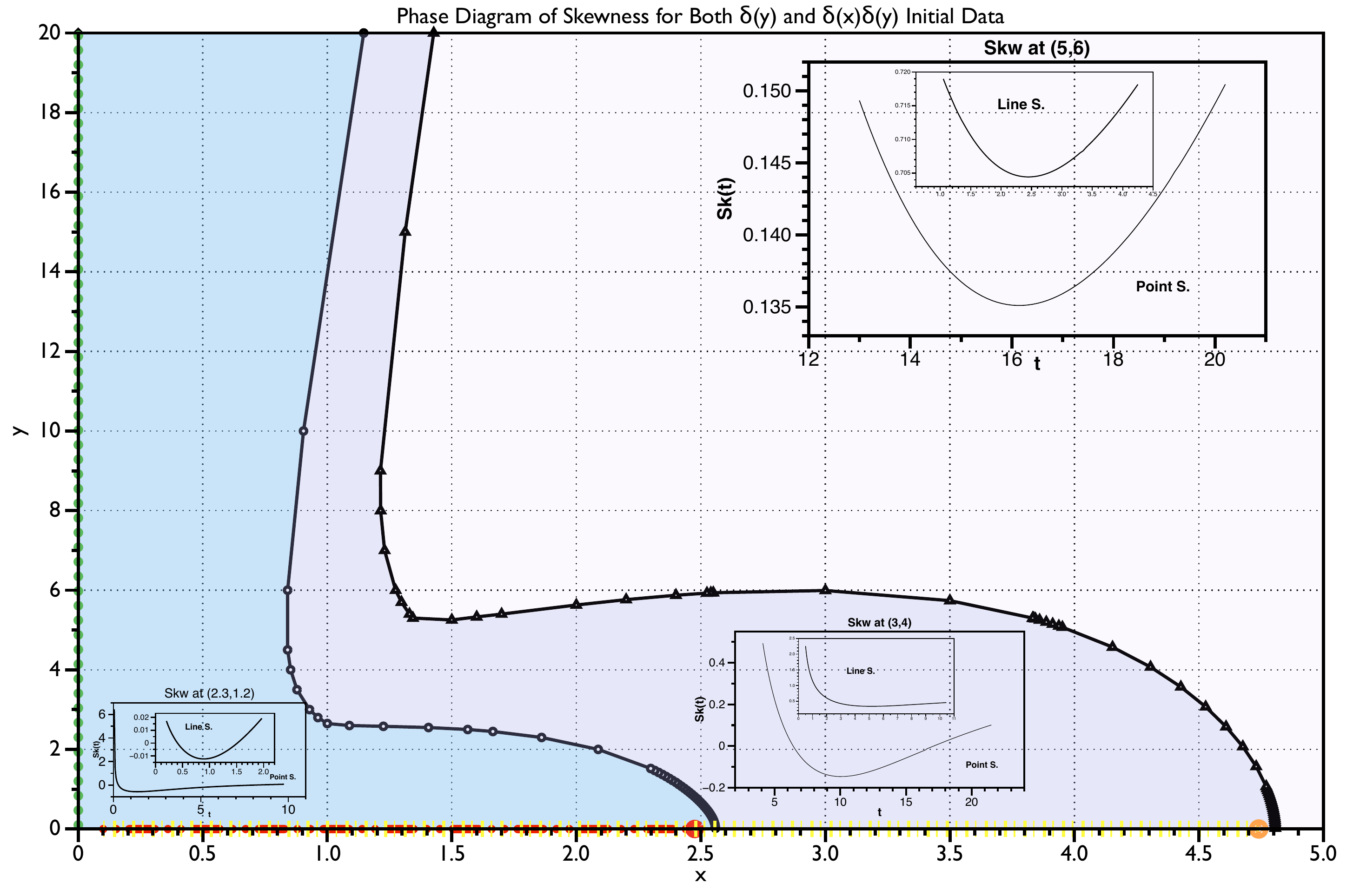}
\caption{{\footnotesize Phase diagram for evolution of statistical skewness in the $x-y$ plane, taking $x_0=0$. Points interior to line source critical (inner) curve experience skewness initially positive and changing signs at least twice before arriving at the long time state, similar for points interior to the point source critical (outer) curve.  Points exterior (to the right) to these curves have skewness positive for all time.  The exception to this behavior is on the $x$-axis where skewness is initially negative for values to the left of the critical points (left point for line source, right point for point source initial data), and for those points, the skewness changes sign only once in time.}} 
\label{phaseD}
\end{figure}

\begin{figure}[h]
\centering
\includegraphics[width=0.8\textwidth]{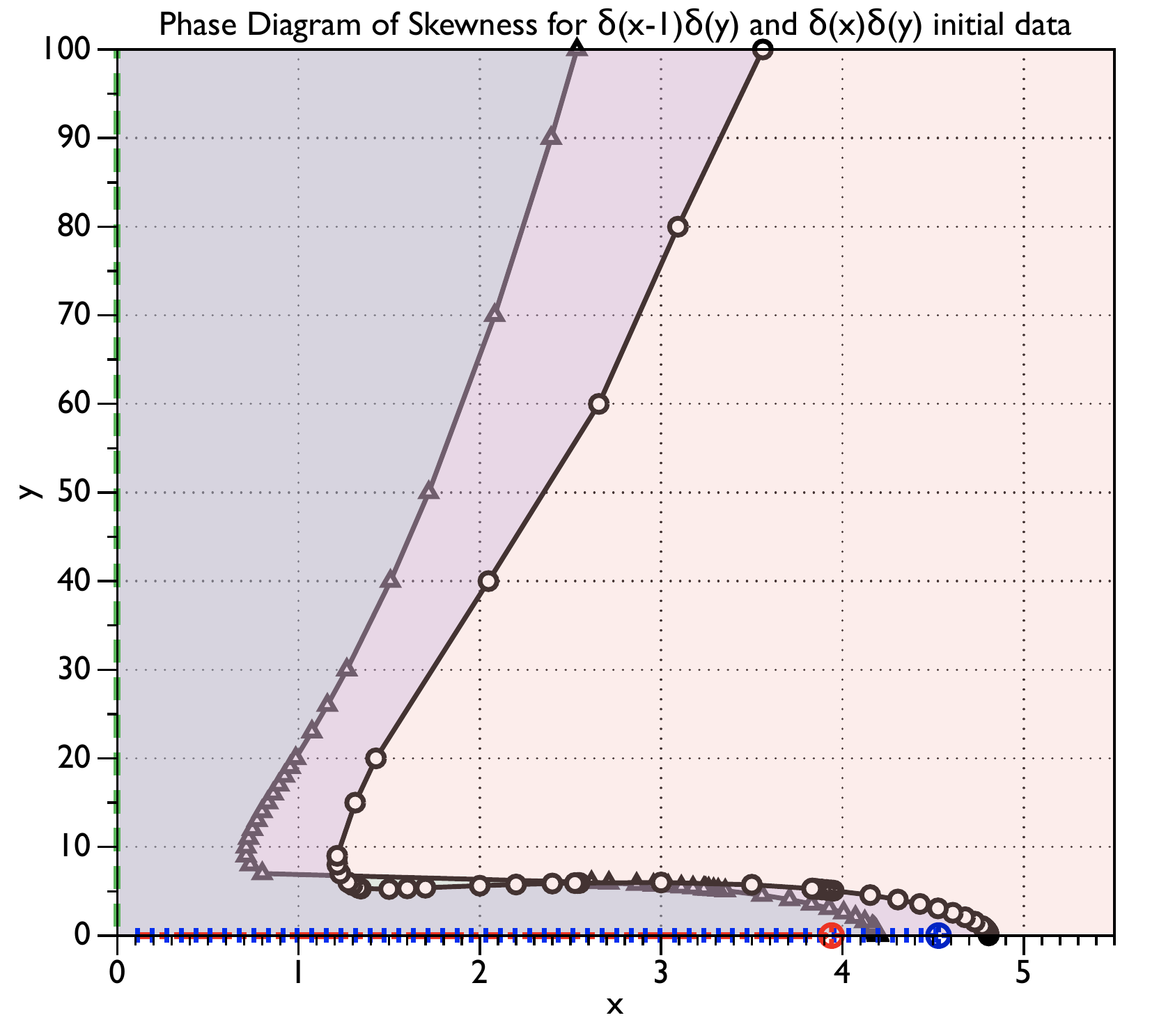}
\caption{{\footnotesize Phase diagram for evolution of statistical skewness in the $x-y$ plane for point source initial data for two different initial locations: $x_0=0, x_0=1$. 
Points interior to the critical curves experience skewness initially positive and changing signs at least twice before arriving at the long time state.  Points exterior (to the right) to these curves have skewness positive for all time.  The exception to this behavior is on the $x$-axis where skewness is initially negative for values to the left of the critical points, and for those points, the skewness changes sign only once in time.}} 
\label{phaseDpoint}
\end{figure}

To compute these regions, we utilize a numerical evaluation of the Mehler kernel integral representation for the centered third moment (given in the appendix), and numerical continuation to identify the critical curves for the line and point source outside of which the skewness is positive for all time (connecting to the asymptotic limit computed in the previous sub-section).  This critical curve is defined by seeking a double zero of the skewness:  given a point $x$, seek the simultaneous solution to the following equations
\begin{eqnarray}
\frac{\partial}{\partial t} S(x,y,t)&=&0\\
S(x,y,t)&=&0
\end{eqnarray}
where $S(x,y,t)$ is the statistical skewness.  The first equation seeks the critical time for which the skewness has a minimum (extrema), while the second condition asks that this minimum value is zero, and this defines a critical curve, $y=y^*(x)$.  For other values of $y$ not on this curve, the skewness will either vanish at least twice in time, or will not vanish.  Shown in Figure \ref{phaseD} is a phase diagram showing these critical curves for the case of the point and line source initial data.  The inset diagrams the local behavior of the skewness evolution at the labeled $(x,y)$ coordinate.  Outside of these curves, the skewness is positive for all time.  Also shown in this figure along the $x$ axis are different segments for the point and line source.  The $x$ axis has peculiarities:  A straightforward short time expansion shows that the zero time limit of the skewness at the origin is $\frac{-8\sqrt{3}}{5}$ for the line source, and for the point source is $\frac{-8\sqrt{5}}{7}$.  This may seem surprising in that the initial data is deterministic; however, since the initial data itself is a delta function, the PDF at t=0 is not well defined since at $t=0$ one is tempted to write, $P(T)=\delta(T-\delta(y))$, and at $y=0$, this is a delta function with support at infinity, a highly singular object.  On the other hand, if the initial condition is taken to be mollified as an initial Gaussian function of $y$, then the short time limit of the skewness is zero.  In any case, for the present discussion of delta function initial data, the skewness will start negative for $x=y=0$, and eventually reach its long time positive limit with a single time change for both the point and line source initial datas.  For points on the $x$ axis away from the origin, this behavior persists, until a critical value of $x=x^*$ is reached.  At this point, the zero time limit of the skewness vanishes.  This point is labeled on the $x$ axis in figure \ref{phaseD}, for the case of line and point source data.  It is interesting that this point lives inside their corresponding critical curve.  Points to the right of this point and to the left of the curve experience a skewness which is initially positive, and changes sign twice to arrive at the positive long time limiting state.  The behavior is different along the $y$-axis:  For any point off the origin on the $y$-axis, the skewness is initially positive and changes sign twice.  The short time asymptotics along this axis are derived in the appendix, and demonstrate this explicit change in behavior for points on the origin to points off the origin on the $y$-axis.  Figure \ref{phaseDpoint} documents similar curves separating regions of the $x-y$ plane for which the skewness is positive for all time, from regions where the skewness changes sign multiple times for point source initial data with $x0=0$, and $x0=1$ to give some insight into the role played by moving the initial point source off the point of zero shear.

We remark that more exotic skewness evolutions may arise through more complex initial conditions.  Having the skewness properties understood for the random Green's function help to interpret what may happen in general.  For example, with multiple line sources properly positioned, the skewness can be observed to change sign many times as pulses from each different line source arrive at a given observation point on different timescales set by a combination of the heat evolution and random advection.  

All of these results are easily observed numerically using a new Monte-Carlo approach, dubbed Direct Monte-Carlo (DMC) developed in the next section.  It will be through the careful benchmarking of this algorithm using the above discussion derived with accurate numerical evaluation of explicit Mehler kernels which provides the ultimate benchmarking of Full Monte-Carlo (FMC) simulations of the underlying stochastic differential equations presented below in section 6 in free space, documenting the numbers of particles and field realizations necessary to replicated the exact Mehler analysis.  With that successful benchmark, the FMC can be applied with vanishing Neumann boundary conditions, easily implemented for a half space or channel domain, and will demonstrate the radically different hot leaning invariant measure.  

\section{A new rapid Monte-Carlo method}

We next develop a rapid Monte-Carlo method built upon using the random Green's function integral representation formulas given in equations (\ref{RGLine}) and (\ref{RGPoint}).  Using the well known time rescaling properties of Brownian motion, the L2 norm of Brownian and the Brownian Bridge in equation (\ref{law}), realizations of the the random passive scalar, for either line source, or point source initial data at any point $(x,y)$ for {\em any time} are obtained through time rescaling. One needs only evaluate accurate realizations of the various stochastic processes at time $t=1$, and then the random Green's function formulae immediate provide realizations of the random field through elementary function evaluation.  In turn, repeated realizations of these processes can be used to construct statistics and PDF histograms at any point in the $x-y$ plane and at any time.  

\begin{figure}[h]
\centering
\includegraphics[width=0.8\textwidth]{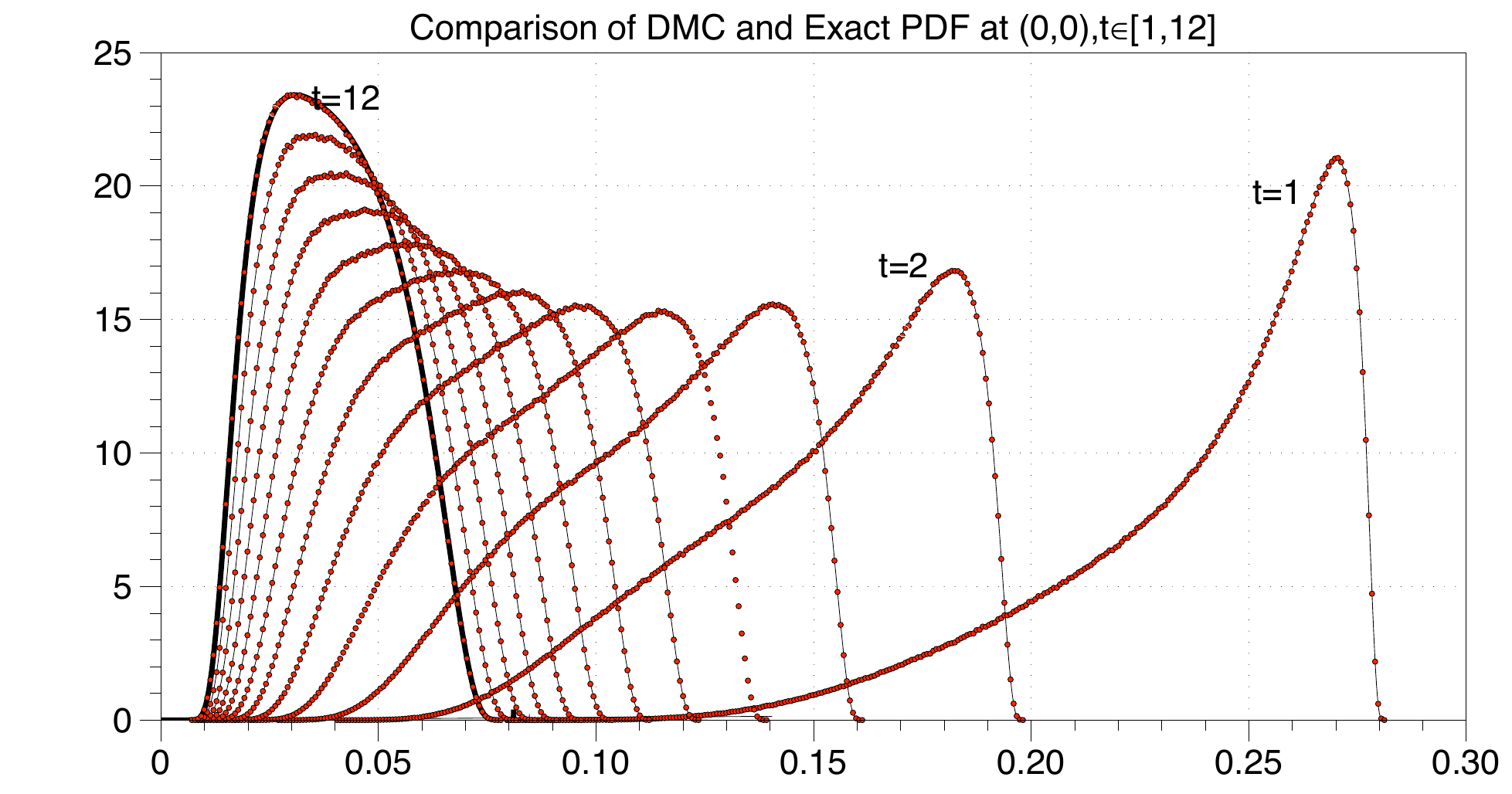}
\label{DMC}
\caption{{\footnotesize Comparison of Direct Monte-Carlo PDF evolution for line source initial data at $x=y=0$ with exact PDF evaluated using equation \ref{loopline}, using $10^8$ realizations of the random process, $\gamma(t)$, and $500$ steps.}} 
\end{figure}

\begin{figure}[h]
\centering
\includegraphics[width=.77\textwidth]{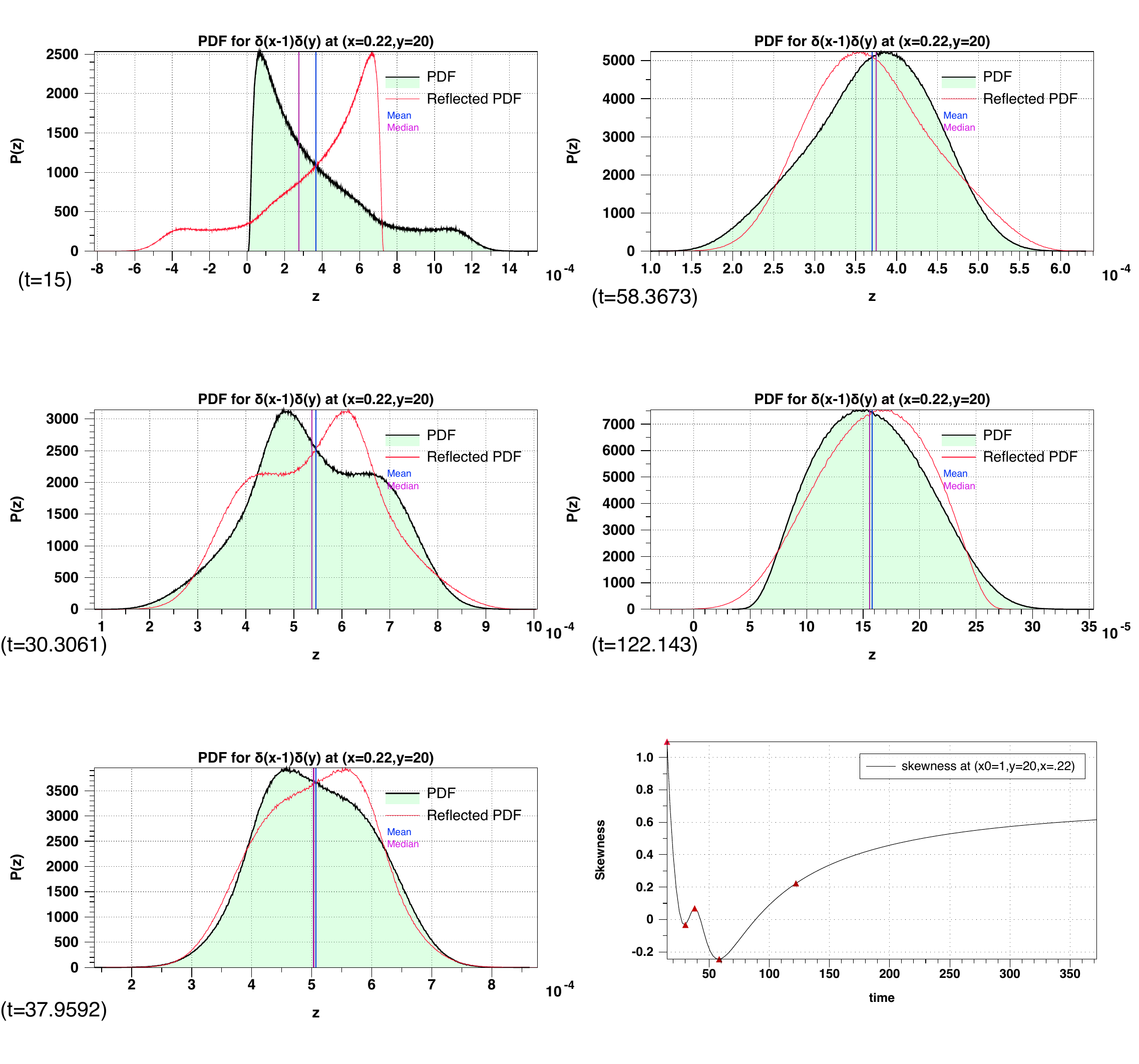}
\label{DMC2}
\caption{{\footnotesize Evolution of the PDF for point source initial data using DMC at $x=0.22,y=20$ and $x0=1$.  Counterclockwise from upper shows the skewness evolution as a function of time, with 4 sign changes at different times:  $t=15$, $t=30.3061$, $t=37.9592$, $t=58.3673$, and $t=122.143$.  Also shown are the means and medians, as well as the PDF reflected about the mean.  Observe, only the case with $t=30.3061$ has multiple crossings to the right of the mean, and hence the statistical significance of the skewness sign change as regards the shaping of the PDF for that time interval is questionable.}} 
\end{figure}

Realizations of the processes $B,\xi,\eta,\mu$ are constructed using pseudo-random number generation and a variation of the Box-Muller transformation (described below in the next section), with integrals of the process evaluated unless otherwise noted using $500$ time steps.  Here, we utilize $10^8$ realizations of each process to construct the various histograms and statistics.  We find with these parameters exceptionally strong agreement with the PDF's developed in the prior section using the exact integration of the loop integrals for the PDF at the origin $x=y=0$ . To demonstrate this success, we show in figure \ref{DMC} the results of our DMC simulation with $x=y=0$ for the line source initial data with results from the exact evaluation from figure \ref{lineevolution} superimposed.  In figure \ref{DMC2}, we show the PDF evolution computed using the DMC algorithm, but with $x=0.22,y=20$ for the point source data taking $x0=1$.   Typically, for points in the interior of the multiple sign change skewness region presented in figure \ref{phaseD}, we see the skewness initially positive, and then experiences two sign changes corresponding to the PDF moving from cold leaning to hot leaning, then back to cold leaning.  However, for the case studied here, we actually see 4 sign changes of the skewness (skewness evolution shown in the bottom right panel of figure \ref{DMC2})!  Counterclockwise from upper left shows the PDF evolution, but also showing each PDFs reflection about its mean for five time values: the first at $t=15$ within the initially positive skewness time interval, the second at $t=30.3061$ within the first interval of negative skewness, thirdly at $t=37.9592$ in the second region of positive skewness, fourthly at $t=58.3673$ in the final region of negative skewness, and lastly at $t=122.143$, in the final interval of positive skewness.  Also shown in each plot is the median of the distribution.  We do this to point out an important feature regarding the sign of the skewness.  In recent work of two of the authors \cite{aminiancamassamcl}, they focussed attention on when the sign of the skewness implies a correlation between the median and the mean.  In particular, if the median is to the right of the mean, more mass is clearly to the right of the mean by definition.  In that work, the authors presented a criterion which guarantees when the sign of the skewness implies such a correlation between the median and the mean.  That criterion is satisfied for PDF distributions whose reflection about the mean possesses a single intersection to the right of the mean with the original PDF.  Inspection of the graphs in figure \ref{DMC2} shows that this single reflection criterion is satisfied for all but the case with $t=30.3061$ in the first time interval of negative skewness.  Clearly, in that case, the median remains to the left of the mean, whereas the skewness is negative, violating the correlation.  But, importantly, there are multiple intersections between the PDF and the reflected PDF about the mean, to the right of the mean.  So the correlation between skewness and loadedness properties is not guaranteed in this instance.  This suggests that the first interval of negative skewness is not physically relevant, but does demonstrate in a physical, non-pathological problem, a case for which a negative sign of the skewness doesn't necessarily imply a hot leaning state.  It is best practice to, when possible plot the evolution of the full PDF.  

\section{Comparison of Full-Monte Carlo Method (FMC) with DMC, and the surprising role of boundaries on the long time state}

We next explore the role of the addition of physical boundary condition in the random passive scalar equation using what we refer to as ``Full Monte-Carlo" simulations (FMC) which refers to solving the underlying stochastic differential equation (SDE) underlying passive scalar advection with respect to the double randomness arising from a) molecular diffusion, and b) the random flow field.  This is in contrast to our new DMC simulations which are built around the free space random Green's function, and their underlying stochastic process time rescaling properties.  FMC is much more expensive than DMC, but necessary to understand the role which boundary conditions play in random PDE, particularly given the lack of translation invariance for both the random Green's functions, as well as the Mehler kernel integral representation formula for the N'th statistical moment.  Given the fact that the work count is at least the square of the work count for DMC, it is necessary to first carefully benchmark the ability of the FMC to faithfully replicated the results of DMC.  This will help to establish numbers of realizations and time step criteria needed to resolve the PDFs and statistics associated with the random PDE with vanishing Neumann boundary conditions.         

To solve the full advection diffusion problem in equation (\ref{passivescalar}), we have 
implemented a Monte-Carlo code in Fortran.  Monte-Carlo methods are advantageous particularly for this class of problems involving complex geometry in that implementing boundary conditions is facilitated through simple billiard like reflection rules.  
The approach is to sample realizations of the 
equivalent stochastic differential equation underlying the advection diffusion equation in non-dimensional form:
\begin{eqnarray*}
dX(\tau) &=&\sqrt{2} dW_1\\
dY(\tau) &=&  u(X(\tau),Z(\tau)) d\tau + \sqrt{2} dW_2 \\
\end{eqnarray*}

For each realization of the random velocity field (typically for the above problem, we take $N_{vel} \approx 10^6$), then our typical initial condition is a delta function which we take to be supported at the center of the channel.  For each realization of the random velocity, each particle is evolved from its initial condition using the above SDE  using $10^6$ independent realizations of the independent Wiener processes $(W_1(t),W_2(t))$, each realization evaluated for $N_t$ time steps (typically $N_t \approx 10^3-10^5$).  This gives an operation count of approximately $10^{15}-10^{17}$ which takes a few days using 200 processors on our super cluster at UNC.  It should be emphasized, that this problem has near perfect parallel scalability as the different random flow realizations (the outer loop) are completely independent and may be run independently on individual processors with no need for node communication at all.  

The white noise increments 
$dW_i$ are standard independent and identically distributed random variables (with reflecting 
boundary conditions for $dW_1$ imposed at $x=0$ and $x=1$). An 
Euler-Maruyama timestepping is used with 
a bounded timestep, typically $\Delta \tau \leq 10^{-3}$. To efficiently evaluate the short time dynamics, we 
use a variable, exponentially growing time step (typically starting from $10^{-8}$, to ensure resolution of the short time evolution), switching to fixed time step when the preset 
threshold bound is reached.  This 
choice strikes a good balance between 
efficiency and accuracy at long time, and 
largely limits the computationally expensive
issue of calculating multiple reflections 
off the boundary during each timestep when 
the domain size is order one. We have verified that this choice agrees with the standard uniform time-stepping approach by extensive comparisons with the exactly solvable cases.
The normal random increments are produced 
through a combination of a Mersenne Twister 
uniform random number generator \cite{matsumoto98} and 
a ``modified polar transformation" method 
(see \cite{marsaglia64}, {\it Introduction}), 
a more efficient variant of the 
well known Box-Muller method for producing 
normally distributed random numbers.

\begin{figure}
\centering
\includegraphics[width=5 in]{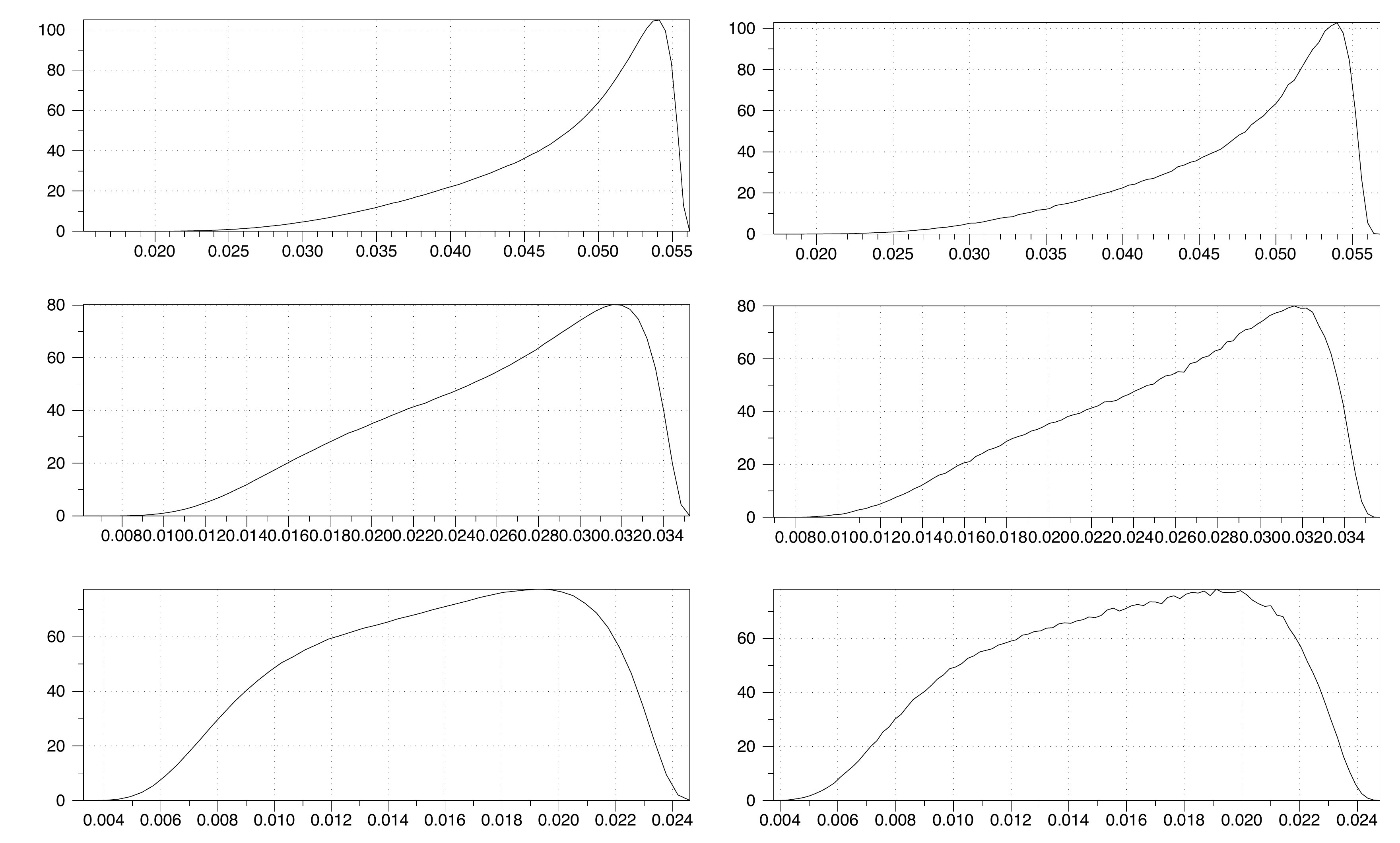}
\caption{{\footnotesize Comparison of Direct Monte-Carlo (left panels) with the Full Monte-Carlo (right panels) at times (t=1,2.5,5) for the random variable, M, with width $a=0.1$, $x0=0$ (both in free space).  Full Monte-Carlo utilizes $10^6$ walkers and $10^6$ random velocity field realizations, with $5*10^3$ total time steps, while the Direct Monte-Carlo has $10^8$ realizations of each underlying Brownian process, with integrals computed with $500$ timesteps.}}  
\label{one}
\end{figure}

To evaluate statistics across many timescales, the Monte-Carlo based methods are advantageous, despite their typically slow convergence rate  
($\simeq 1/\sqrt{N}$), since they are ``gridless,"  and the SDE approach 
directly matches the microscopic physics (compared to 
finite difference/element approaches). 
These methods have been strongly benchmarked for the exactly solvable 
case of laminar channel flow \cite{PRL}, as well as the circular pipe \cite{Barton}, and with these 
parameters for the so called Peclet number, $\Pe \leq 10^4$ 
through the diffusive timescale, the simulation for the spatial skewness has absolute error bounded by $10^{-4}$, in accordance with the law of large numbers.
Similar convergence studies have been applied to the circular pipe, with 
similar results.

Given this Full Monte-Carlo algorithm, we may turn to assess its success in predicting the scalar statistics inherited from the random fluid flow.  The fundamental question to be next assessed is if this procedure has sufficient numbers of flow realizations to accurately capture the statistics of the random passive scalar equation.  Armed with the combination of exact Mehler analysis, our new Direct Monte-Carlo simulations, and new measure changed exact PDF formulae developed above, we next establish the strong quantitative capabilities of this approach for the case of the white in time random linear shear layer in free space.  To this end, a natural random variable to consider is the number of particles in an infinite strip in the $x$-direction, and of width $2a$ in the $y$-direction.  An explicit random representation for realizations of this random variable is directly available through the random Green's function:

\begin{eqnarray}
M &=&\int_{-a}^{a} dy\int_{\mathbb{R}}{dxT(t,x,y,x_{0})}\\
&=&\frac{1}{2} \left(\textrm{Erf}\left(\frac{\sqrt{t}B(1) \textit{Pe}  x_{0}+a}{ \sqrt{4t
   \left(1+t\textit{Pe}^2  \left(\mu+(\beta -B(1))^2
   \right)\right)}}\right)\right) \nonumber \\
   &&-\frac{1}{2}\left(\textrm{Erf}\left(\frac{\sqrt{t}B(1) \textit{Pe}  x_0-a}{
   \sqrt{4t \left(1+t\textit{Pe}^2  \left(\mu+ (\beta -B(1))^2 \right)\right)}}\right)\right)
\end{eqnarray}


We may now directly compare the results of a Full Monte-Carlo simulation with the Direct Monte-Carlo simulation.  Figure \ref{one} depicts this comparison for a point source released at the origin at 3 output times $t=1.0,2.5,5.0$.  The left panels show the result of the Direct Monte-Carlo, while right panels show the result of the ``Full-Monte".  See caption for additional details.  The agreement is clearly excellent, and confirms the ability of the Full Monte-Carlo approach to accurately track the evolving PDFs.  
\begin{figure}
\centering
\includegraphics[width=4.8 in]{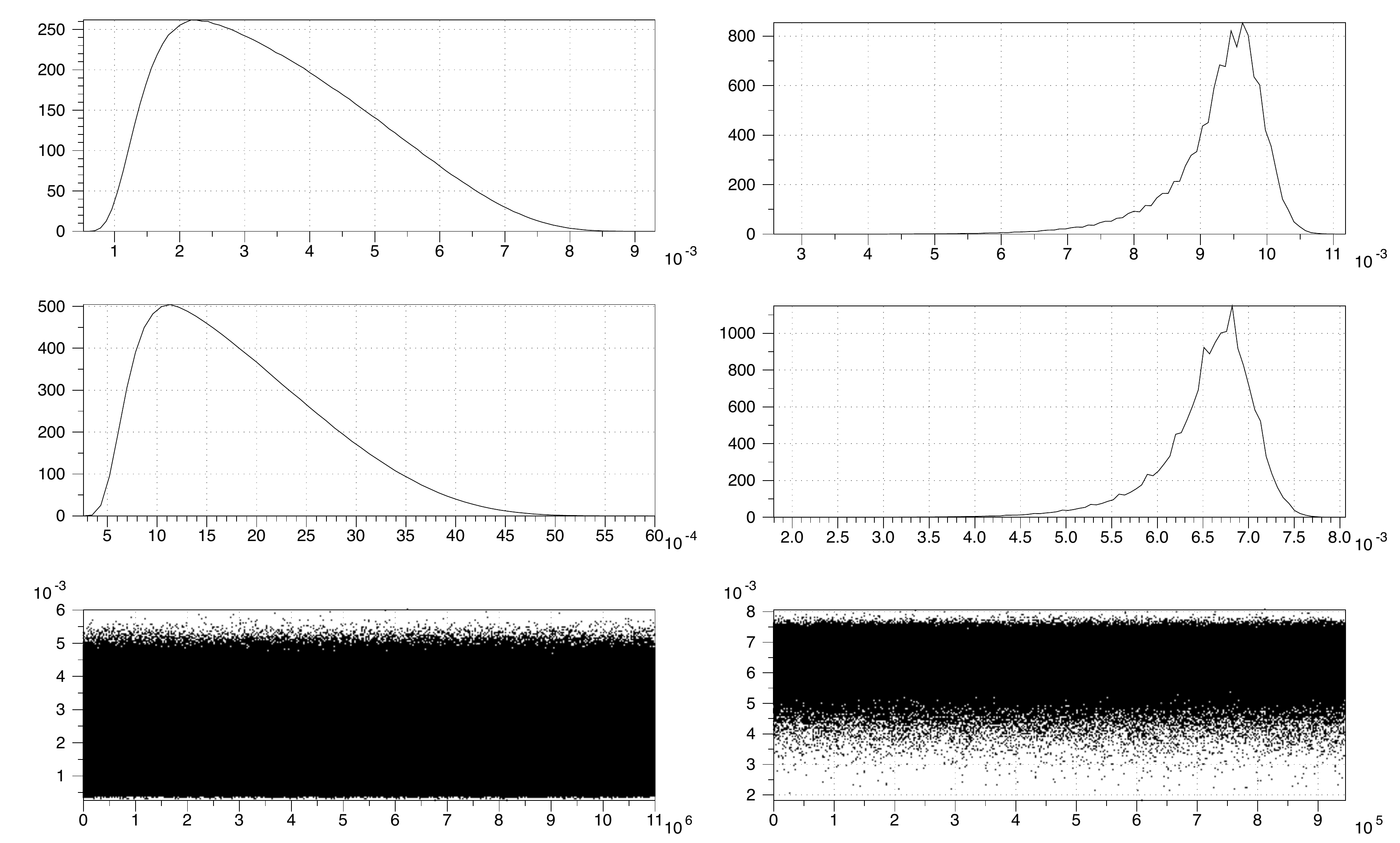}
\caption{{\footnotesize Top panels: Comparison of PDFs Direct Monte-Carlo (left panels) in free space, with the channel Full Monte-Carlo (right panels) at times (t=30,60) for the random variable, $M$, with width $a=0.1$, $x0=1/2$ with no-flux boundary conditions (reflecting) at $x=0$ and $x=1$ .  Full Monte-Carlo utilizes $10^5$ walkers and $10^6$ random velocity field realizations, with $800$ total time steps, while the Direct Monte-Carlo has $10^8$ realizations of each underlying Brownian process, with integrals computed with $500$ timesteps.  Observe the stark difference between the cold-leaning free space calculation, and the hot-leaning, channel simulation.  Bottom panel, realizations of the random variable $M$ in free-space (left), and channel domain (right) at time $t=60$.}}
\label{FullMonty}
\end{figure}

We next present the Monte-Carlo simulations for a particle released at $x=1/2$, for a bounded domain with vanishing Neumann conditions (reflecting) imposed at $x=0$ and $x=1$, here with Peclet number $Pe=1$.  In figure \ref{FullMonty} we show the PDF distributions in the left panels for free space, and in the right panels for the channel, at times $t=30$, and $t=60$.  Clearly, the simulation in the bounded channel is experiencing an entirely different relaxation to its vanishing state than in free space, with its distribution being heavily biased towards the hot state.  In the bottom panels, we show the realizations of the random variable, $M$, for free space (on the left), and for the channel domain (on the right), clearly showing the highly different bias, with the free-space fluctuations biased towards the cold state, while the channel domain is heavily biased towards the hot state.  We remark that this behavior persists well beyond the diffusion timescale in the channel domain, and hence, it is expected to persist for all time.

\section{Concluding remarks} 

Here we have explored the difference between the free space behavior of the Majda model, and enforcing zero flux boundary conditions on an infinite channel.  Using a combination of exact analysis, careful evaluation of Mehler Kernels, and a new rapid Monte-Carlo algorithm (DMC) built upon the random Green's function, we have presented a thorough study of the evolution of the scalar PDF in the $x-y$ plane on short, intermediate, and long time scales, and have established regions in the plane with skewness changing signs multiple times corresponding to exchanges in symmetry between hot and cold leaning states along the way to the eventual long time, limiting, invariant collapsing cold leaning measure.  This DMC provides a strict benchmark which demonstrates criteria for full Monte-Carlo simulations (FMC) of the underlying stochastic differential equations, looping over both random velocity fields, and random walks to faithfully match the exact results.  Armed with this benchmark, FMC is applied to the Majda model with no-flux boundary conditions applied to the walls of an infinite channel, and we observe that the long time invariant measure is seen to collapse with a {\em negative} skewness, corresponding to a hot leaning state.  

Future directions for this work include the development of a physical experiment using a robotically actuated wall to set up simulated random shear layers as well as using random pressure driven flows.  Naturally, it will be interesting to extend these results to non-white in time statistics, particularly those with finite correlation times, and to flows arising in more exotic cross-sectional domains.  Lastly, we remark that rigorous analysis may be applied to compute the long time asymptotics of the statistical skewness in the presence of no-flux boundaries using the method employed by Bronski and McLaughlin \cite{BronskiMcL1}, however that analysis utilized random symmetric initial data, and only was carried out to second order in the coupling coefficient.  In the present discussion, the analysis {\em must} be carried out at least to third order to develop a proper theory for measuring the sign of the statistical skewness at long times.  This calculation is being carried out and will be reported soon.
 
\section{Acknowledgements} 

Camassa and McLaughlin acknowledge funding from the Office of Naval Research (grant
DURIP N00014-12-1-0749) and the NSF (grants RTG
DMS-0943851, CMG ARC-1025523, DMS-1009750, and
DMS-1517879), and Kilic was partially supported through SAMSI graduate fellowship.

\appendix
\section{ Appendix:  Moment Formulae and asymptotic expansions}
\setcounter{equation}{1}

Here we summarize the analysis of the various Mehler kernel integral representations for the statistical moments considered in this paper.  
These formula follow from the work of Majda \cite{Majda}, McLaughlin and Majda \cite{McLMajda}, and dissertation of Kilic \cite{Kilic}.

\begin{footnotesize}
\subsection{Line Initial Data}
First, we present the moment formulae for the case of line source initial data:
\subsubsection{First Moment}

\begin{eqnarray*}
\langle T\rangle =\int_{{{R}}}{dk \exp\left(2\pi i k y\right)\exp\left(-4\pi^{2}k^{2}t\right)\frac{\exp\left(-\frac{\sqrt{2\pi^{2}|k|^{2}}}{2}x^{2}\tanh\left(2\sqrt{2}\pi|k|t\right)\right)}{\sqrt{\cosh\left(2\sqrt{2}\pi |k|t\right)}}}
\end{eqnarray*}

\subsubsection{Second Moment}

\begin{eqnarray*}\langle T^{2}\rangle&=\int_{{{R^{2}}}}{dk_{1}dk_{2} \exp\left(2\pi i k_{1} y\right)\exp\left(2\pi i k_{2} y\right)\exp\left(-4\pi^{2}|\vec{k}|^{2}t\right)\frac{\exp\left(-\frac{\sqrt{2\pi^{2}|\vec{k}|^{2}}}{2}\left
(\frac{k_{1}x_{1}+k_{2}x_{2}}{\sqrt{k_{1}^{2}+k_{2}^{2}}}\right)^{2}\tanh\left(2\sqrt{2}\pi|\vec{k}|t\right)\right)}{\sqrt{\cosh\left(2\sqrt{2}\pi |\vec{k}|t\right)}}}\\
&=\int_{0}^{\infty}\int_{R}{dk_{1}dk_{2}\exp\left(2\pi ik_{1}y\right)\exp\left(2\pi ik_{2}y\right)\exp\left(-4\pi^{2}|\vec{k}|^{2}t\right)\frac{\exp\left(-\frac{\sqrt{2\pi^{2}|\vec{k}|^{2}}}{2}\left(\frac{k_{1}x_{1}+k_{2}x_{2}}{|\vec{k}|}\right)^{2}\tanh\left(2\sqrt{2}\pi|\vec{k}|t\right)\right)}{\sqrt{\cosh\left(2\sqrt{2}\pi |\vec{k}|t\right)}}}\\
&+\int_{0}^{\infty}\int_{R}{dk_{1}dk_{2}\exp\left(-2\pi ik_{1}y\right)\exp\left(2\pi ik_{2}y\right)\exp\left(-4\pi^{2}|\vec{k}|^{2}t\right)\frac{\exp\left(-\frac{\sqrt{2\pi^{2}|\vec{k}|^{2}}}{2}\left(\frac{-k_{1}x_{1}+k_{2}x_{2}}{|\vec{k}|}\right)^{2}\tanh\left(2\sqrt{2}\pi|\vec{k}|t\right)\right)}{\sqrt{\cosh\left(2\sqrt{2}\pi |\vec{k}|t\right)}}}
\end{eqnarray*}

After setting $x_{1}=x_{2}=x,$
\begin{eqnarray*}
\langle T^{2}\rangle=&\int_{0}^{\infty}\int_{0}^{\infty}{dk_{1}dk_{2}\exp\left(2\pi i k_{1}y\right)\exp\left(2\pi i k_{2}y\right)\exp\left(-4\pi^{2}|\vec{k}|^{2}t\right)\frac{\exp\left(\frac{\sqrt{2\pi^{2}|\vec{k}|^{2}}}{2}\left(\frac{k_{1}+k_{2}}{|\vec{k}|}\right)^{2}x^{2}\tanh\left(2\sqrt{2}\pi |\vec{k}|t\right)\right)}{\sqrt{\cosh\left(2\sqrt{2}\pi |\vec{k}|t\right)}}}\\
+&\int_{0}^{\infty}\int_{R}{dk_{1}dk_{2}\exp\left(2\pi ik_{1}y\right)\exp\left(-2\pi ik_{2}y\right)\exp\left(-4\pi^{2}|\vec{k}|^{2}t\right)\frac{\exp\left(\frac{\sqrt{2\pi^{2}|\vec{k}|^{2}}}{2}\left(\frac{k_{1}-k_{2}}{|\vec{k}|}\right)^{2}x^{2}\tanh\left(2\sqrt{2}\pi |\vec{k}|t\right)\right)}{\sqrt{\cosh\left(2\sqrt{2}\pi |\vec{k}|t\right)}}}\\
+&\int_{0}^{\infty}\int_{R}{dk_{1}dk_{2}\exp\left(-2\pi ik_{1}y\right)\exp\left(2\pi ik_{2}y\right)\exp\left(-4\pi^{2}|\vec{k}|^{2}t\right)\frac{\exp\left(\frac{\sqrt{2\pi^{2}|\vec{k}|^{2}}}{2}\left(\frac{-k_{1}+k_{2}}{|\vec{k}|}\right)^{2}x^{2}\tanh\left(2\sqrt{2}\pi |\vec{k}|t\right)\right)}{\sqrt{\cosh\left(2\sqrt{2}\pi |\vec{k}|t\right)}}}\\
+&\int_{0}^{\infty}\int_{R}{dk_{1}dk_{2}\exp\left(-2\pi ik_{1}y\right)\exp\left(-2\pi ik_{2}y\right)\exp\left(-4\pi^{2}|\vec{k}|^{2}t\right)\frac{\exp\left(\frac{\sqrt{2\pi^{2}|\vec{k}|^{2}}}{2}\left(\frac{-k_{1}-k_{2}}{|\vec{k}|}\right)^{2}x^{2}\tanh\left(2\sqrt{2}\pi |\vec{k}|t\right)\right)}{\sqrt{\cosh\left(2\sqrt{2}\pi |\vec{k}|t\right)}}}
\end{eqnarray*}
\begin{eqnarray*}
\langle T^{2}\rangle=&2\int_{0}^{\infty}\int_{0}^{\infty}{dk_{1}dk_{2}\exp\left(2\pi i k_{1}y\right)\exp\left(2\pi i k_{2}y\right)\exp\left(-4\pi^{2}|\vec{k}|^{2}t\right)\frac{\exp\left(\frac{\sqrt{2\pi^{2}|\vec{k}|^{2}}}{2}\left(\frac{k_{1}-k_{2}}{|\vec{k}|}\right)^{2}x^{2}\tanh\left(2\sqrt{2}\pi |\vec{k}|t\right)\right)}{\sqrt{\cosh\left(2\sqrt{2}\pi |\vec{k}|t\right)}}}\\
+&2\int_{0}^{\infty}\int_{R}{dk_{1}dk_{2}\exp\left(2\pi ik_{1}y\right)\exp\left(-2\pi ik_{2}y\right)\exp\left(-4\pi^{2}|\vec{k}|^{2}t\right)\frac{\exp\left(\frac{\sqrt{2\pi^{2}|\vec{k}|^{2}}}{2}\left(\frac{k_{1}+k_{2}}{|\vec{k}|}\right)^{2}x^{2}\tanh\left(2\sqrt{2}\pi |\vec{k}|t\right)\right)}{\sqrt{\cosh\left(2\sqrt{2}\pi |\vec{k}|t\right)}}}
\end{eqnarray*}
\begin{eqnarray*}
\langle T^{2}\rangle=&2\int_{0}^{\infty}\int_{0}^{\infty}{dk_{1}dk_{2}\cos\left(2\pi \left(k_{1}+k_{2}\right)y\right)\exp\left(-4\pi^{2}|\vec{k}|^{2}t\right)\frac{\exp\left(\frac{\sqrt{2\pi^{2}|\vec{k}|^{2}}}{2}\left(\frac{k_{1}+k_{2}}{|\vec{k}|}\right)^{2}x^{2}\tanh\left(2\sqrt{2}\pi |\vec{k}|t\right)\right)}{\sqrt{\cosh\left(2\sqrt{2}\pi |\vec{k}|t\right)}}}\\
+&2\int_{0}^{\infty}\int_{0}^{\infty}{dk_{1}dk_{2}\cos\left(2\pi \left(k_{1}-k_{2}\right)y\right)\exp\left(-4\pi^{2}|\vec{k}|^{2}t\right)\frac{\exp\left(\frac{\sqrt{2\pi^{2}|\vec{k}|^{2}}}{2}\left(\frac{k_{1}-k_{2}}{|\vec{k}|}\right)^{2}x^{2}\tanh\left(2\sqrt{2}\pi |\vec{k}|t\right)\right)}{\sqrt{\cosh\left(2\sqrt{2}\pi |\vec{k}|t\right)}}}
\end{eqnarray*}
If we set $k_{1}=r\cos\left(\phi\right),$ and $k_{2}=r\sin\left(\phi\right),$
\begin{eqnarray*}
\langle T^{2}\rangle=&2\int_{0}^{\infty}\int_{0}^{\frac{\pi}{2}}{drd\phi\cos\left(2\pi r \left(\cos(\phi)+\sin(\phi)\right)y\right)\exp\left(-4\pi^{2}r^{2}t\right)\frac{\exp\left(\frac{\sqrt{2\pi^{2}}r}{2}\left({\cos(\phi)+\sin(\phi)}\right)^{2}x^{2}\tanh\left(2\sqrt{2}\pi rt\right)\right)}{\sqrt{\cosh\left(2\sqrt{2}\pi rt\right)}}}\\
+&2\int_{0}^{\infty}\int_{0}^{\frac{\pi}{2}}{dk_{1}dk_{2}\cos\left(2\pi r\left(\cos(\phi)-\sin(\phi)\right)y\right)\exp\left(-4\pi^{2}r^{2}t\right)\frac{\exp\left(\frac{\sqrt{2\pi^{2}}r}{2}\left({\cos(\phi)-\sin(\phi)}\right)^{2}x^{2}\tanh\left(2\sqrt{2}\pi rt\right)\right)}{\sqrt{\cosh\left(2\sqrt{2}\pi rt\right)}}}
\end{eqnarray*}

\subsubsection{Third Moment}
\begin{eqnarray*}
\langle T^{3}\rangle=&\int_{{{R^{3}}}}{dk_{1}dk_{2}dk_{3} \exp\left(2\pi i k_{1} y\right)\exp\left(2\pi i k_{2} y\right)\exp\left(2\pi i k_{3} y\right)}\\
&{\exp\left(-4\pi^{2}|\vec{k}|^{2}t\right)\frac{\exp\left(-\frac{\sqrt{2\pi^{2}|\vec{k}|^{2}}}{2}\left
(\frac{k_{1}x_{1}+k_{2}x_{2}+k_{3}x_{3}}{\sqrt{k_{1}^{2}+k_{2}^{2}+k_{3}^{2}}}\right)^{2}\tanh\left(2\sqrt{2}\pi|\vec{k}|t\right)\right)}{\sqrt{\cosh\left(2\sqrt{2}\pi |\vec{k}|t\right)}}}
\end{eqnarray*}
\begin{eqnarray*}
\langle T^{3}\rangle=&2\int_{0}^{\infty}\int_{0}^{\infty}\int_{0}^{\infty}{dk_{1}dk_{2}dk_{3}\cos\left(2\pi \left(k_{1}+k_{2}+k_{3}\right)y\right)\exp\left(-4\pi^{2}|\vec{k}|^{2}t\right)\frac{\exp\left(\frac{\sqrt{2\pi^{2}|\vec{k}|^{2}}}{2}\left(\frac{k_{1}+k_{2}+k_{3}}{|\vec{k}|}\right)^{2}x^{2}\tanh\left(2\sqrt{2}\pi |\vec{k}|t\right)\right)}{\sqrt{\cosh\left(2\sqrt{2}\pi |\vec{k}|t\right)}}}\\
+&2\int_{0}^{\infty}\int_{0}^{\infty}\int_{0}^{\infty}{dk_{1}dk_{2}dk_{3}\cos\left(2\pi \left(k_{1}+k_{2}-k_{3}\right)y\right)\exp\left(-4\pi^{2}|\vec{k}|^{2}t\right)\frac{\exp\left(\frac{\sqrt{2\pi^{2}|\vec{k}|^{2}}}{2}\left(\frac{k_{1}+k_{2}-k_{3}}{|\vec{k}|}\right)^{2}x^{2}\tanh\left(2\sqrt{2}\pi |\vec{k}|t\right)\right)}{\sqrt{\cosh\left(2\sqrt{2}\pi |\vec{k}|t\right)}}}\\
+&2\int_{0}^{\infty}\int_{0}^{\infty}\int_{0}^{\infty}{dk_{1}dk_{2}dk_{3}\cos\left(2\pi \left(k_{1}-k_{2}+k_{3}\right)y\right)\exp\left(-4\pi^{2}|\vec{k}|^{2}t\right)\frac{\exp\left(\frac{\sqrt{2\pi^{2}|\vec{k}|^{2}}}{2}\left(\frac{k_{1}-k_{2}+k_{3}}{|\vec{k}|}\right)^{2}x^{2}\tanh\left(2\sqrt{2}\pi |\vec{k}|t\right)\right)}{\sqrt{\cosh\left(2\sqrt{2}\pi |\vec{k}|t\right)}}}\\
+&2\int_{0}^{\infty}\int_{0}^{\infty}\int_{0}^{\infty}{dk_{1}dk_{2}dk_{3}\cos\left(2\pi \left(k_{1}-k_{2}-k_{3}\right)y\right)\exp\left(-4\pi^{2}|\vec{k}|^{2}t\right)\frac{\exp\left(\frac{\sqrt{2\pi^{2}|\vec{k}|^{2}}}{2}\left(\frac{k_{1}-k_{2}-k_{3}}{|\vec{k}|}\right)^{2}x^{2}\tanh\left(2\sqrt{2}\pi |\vec{k}|t\right)\right)}{\sqrt{\cosh\left(2\sqrt{2}\pi |\vec{k}|t\right)}}}
\end{eqnarray*}
$k_{1}=r\sin(\phi)\cos(\theta),$$k_{2}=r\sin(\phi)\sin(\theta),$$k_{3}=r\cos(\phi)$ for $0\leq \theta \leq \frac{\pi}{2},$ $0\leq \phi \leq \frac{\pi}{2},$ $0<r<\infty.$ 
\subsection{Point Source}
Next, we present the moment formulae for point source initial data:

\subsubsection{Mean}
\begin{eqnarray*}
&&\langle T\rangle=2\int_{0}^{\infty}dr\frac{\exp\left(-4\pi^{2}r^{2}t\right){\sqrt{\sqrt{2}\pi r}}}{\sqrt{2\pi\sinh\left(2\sqrt{2}\pi r t\right)}}\cos\left(2\pi y r\right)\exp\left(\frac{\pi r}{\sqrt{2}}\coth\left(2\sqrt{2}\pi r t\right)\left(x^{2}+x_{0}^{2}\right)\right)\\
&&\exp\left(\frac{\sqrt{2}\pi r xx_{0}}{\sinh\left(2\sqrt{2}\pi r t\right)}\right)
\end{eqnarray*}
\subsubsection{Second Field Moment}
\begin{eqnarray*}
\langle T^{2}\rangle&=2\int_{0}^{\infty}\int_{0}^{\frac{\pi}{2}}{drd\phi\cos\left(2\pi r \left(\cos\left(\phi\right)+\sin\left(\phi\right)\right)y\right)\exp\left(-4\pi^{2}r^{2}t\right)\frac{\exp\left(\frac{\sqrt{2\pi^{2}r^{2}}}{2}\left(1+\sin\left(2\phi\right)\right)\left(x^{2}+x_{0}^{2}\right)\coth\left(2\sqrt{2}\pi rt\right)\right)}{\sqrt{(4\pi t)2\pi \sinh\left(2\sqrt{2}\pi rt\right)}}}\\
&{{\left|\left(2\pi^{2}r^{2}\right)^{\frac{1}{4}}\right|}\exp\left(-\frac{(x-x_{0})^{2}}{4t}\left(1-\sin\left(2\phi\right)\right)\right)}\exp\left(\frac{xx_{0}\left(1+\sin(2\phi)\right)\sqrt{2}\pi r}{\sinh{\left(2\sqrt{2}\pi r t\right)}}\right)\\
&+2\int_{0}^{\infty}\int_{0}^{\frac{\pi}{2}}{drd\phi\cos\left(2\pi r\left(\cos\left(\phi\right)-\sin\left(\phi\right)\right)y\right)\exp\left(-4\pi^{2}r^{2}t\right)\frac{\exp\left(\frac{\sqrt{2\pi^{2}r^{2}}}{2}\left(1-\sin\left(2\phi\right)\right)(x^{2}+x_{0}^{2})\coth\left(2\sqrt{2}\pi rt\right)\right)}{\sqrt{(4\pi t)2\pi\sinh\left(2\sqrt{2}\pi rt\right)}}}\\
&{{\left|\left(2\pi^{2}r^{2}\right)^{\frac{1}{4}}\right|}\exp\left(-\frac{(x-x_{0})^{2}}{4t}\left(1+\sin\left(2\phi\right)\right)\right)}\exp\left(\frac{xx_{0}\left(1-\sin(2\phi)\right)\sqrt{2}\pi r}{\sinh{\left(2\sqrt{2}\pi r t\right)}}\right)
\end{eqnarray*}
\subsubsection{Third Field Moment}
\begin{eqnarray*}
&&\langle T^{3}\rangle= 2\int_{0}^{\infty}\int_{0}^{\frac{\pi}{2}}\int_{0}^{\frac{\pi}{2}}{drd\phi d\theta}{r^{2}\sin\left(\phi \right)\cos\left[2\pi r\left[\sin\left(\phi\right)\left(\cos\left(\theta\right)+\sin\left(\theta\right)\right)+\cos\left(\phi\right)\right]y\right]}\\
&&\exp\left(\frac{\left(\pi r\right)}{\sqrt{2}}\left(\sin\left(\phi\right)\left(\sin(\theta)+\cos(\theta)\right)+\cos(\phi)\right)(x^{2}+x_{0}^{2})\coth\left(2\sqrt{2}\pi r t\right)\right)\\
&&\frac{\exp\left(-4\pi^{2}r^{2}t\right)|(2\pi^{2}r^{2})^{\frac{1}{4}}|}{\sqrt{\left(4\pi t\right)^{2}2\pi\sinh\left(2\sqrt{2}\pi rt\right)}}\exp\left(-\frac{(x-x_{0})^{2}}{4t}\left(1-\sin(2\theta)\right)\right)\\
&&\exp\left(\frac{xx_{0}\left(\sin\left(\phi\right)\left(\cos\left(\theta\right)+\sin\left(\theta\right)\right)+\cos\left(\phi\right))\right)\sqrt{2}\pi r}{\sinh{\left(2\sqrt{2}\pi r t\right)}}\right)
\\
&&\exp\left(-\frac{(x-x_{0})^{2}}{4t}\left(\frac{-\left(\sin\left(\theta\right)+\cos\left(\theta\right)\right)\cos\left(\phi\right)+\sin\left(\phi\right)}{r\cos\left(\phi\right)}\right)^{2}\right)\\
&&+2\int_{0}^{\infty}\int_{0}^{\frac{\pi}{2}}\int_{0}^{\frac{\pi}{2}}{drd\phi d\theta}{r^{2}\sin\left(\phi \right)\cos\left[2\pi r\left[\sin\left(\phi\right)\left(\cos\left(\theta\right)+\sin\left(\theta\right)\right)-\cos\left(\phi\right)\right]y\right]}\\
&&\exp\left(\frac{\left(\pi r\right)}{\sqrt{2}}\left(\sin\left(\phi\right)\left(\sin(\theta)+\cos(\theta)\right)-\cos(\phi)\right)(x^{2}+x_{0}^2)\coth\left(2\sqrt{2}\pi r t\right)\right)\\
&&\frac{\exp\left(-4\pi^{2}r^{2}t\right)|(2\pi^{2}r^{2})^{\frac{1}{4}}|}{\sqrt{\left(4\pi t\right)^{2}2\pi\sinh\left(2\sqrt{2}\pi rt\right)}}\exp\left(-\frac{(x-x_{0})^{2}}{4t}\left(1-\sin(2\theta)\right)\right)\\
&&\exp\left(\frac{xx_{0}\left(\sin\left(\phi\right)\left(\cos\left(\theta\right)+\sin\left(\theta\right)\right)-\cos\left(\phi\right))\right)\sqrt{2}\pi r}{\sinh{\left(2\sqrt{2}\pi r t\right)}}\right)\\
&&\exp\left(-\frac{(x-x_{0})^{2}}{4t}\left(\frac{-\left(\sin\left(\theta\right)+\cos\left(\theta\right)\right)\cos\left(\phi\right)-\sin\left(\phi\right)}{r\cos\left(\phi\right)}\right)^{2}\right)\\
&&+2\int_{0}^{\infty}\int_{0}^{\frac{\pi}{2}}\int_{0}^{\frac{\pi}{2}}{drd\phi d\theta}{r^{2}\sin\left(\phi \right)\cos\left[2\pi r\left[\sin\left(\phi\right)\left(\cos\left(\theta\right)-\sin\left(\theta\right)\right)+\cos\left(\phi\right)\right]y\right]}\\
&&\exp\left(\frac{\left(\pi r\right)}{\sqrt{2}}\left(\sin\left(\phi\right)\left(-\sin(\theta)+\cos(\theta)\right)+\cos(\phi)\right)(x^{2}+x_{0}^2)\coth\left(2\sqrt{2}\pi r t\right)\right)\\
&&\frac{\exp\left(-4\pi^{2}r^{2}t\right)|(2\pi^{2}r^{2})^{\frac{1}{4}}|}{\sqrt{\left(4\pi t\right)^{2}2\pi\sinh\left(2\sqrt{2}\pi rt\right)}}\exp\left(-\frac{(x-x_{0})^{2}}{4t}\left(1+\sin(2\theta)\right)\right)\\
&&\exp\left(\frac{xx_{0}\left(\sin\left(\phi\right)\left(\cos\left(\theta\right)-\sin\left(\theta\right)\right)+\cos\left(\phi\right))\right)\sqrt{2}\pi r}{\sinh{\left(2\sqrt{2}\pi r t\right)}}\right)\\
&&\exp\left(-\frac{(x-x_{0})^{2}}{4t}\left(\frac{\left(\sin\left(\theta\right)-\cos\left(\theta\right)\right)\cos\left(\phi\right)+\sin\left(\phi\right)}{r\cos\left(\phi\right)}\right)^{2}\right)\\
&&+2\int_{0}^{\infty}\int_{0}^{\frac{\pi}{2}}\int_{0}^{\frac{\pi}{2}}{drd\phi d\theta}{r^{2}\sin\left(\phi \right)\cos\left[2\pi r\left[\sin\left(\phi\right)\left(\cos\left(\theta\right)-\sin\left(\theta\right)\right)-\cos\left(\phi\right)\right]y\right]}\\
&&\exp\left(\frac{\left(\pi r\right)}{\sqrt{2}}\left(\sin\left(\phi\right)\left(-\sin(\theta)+\cos(\theta)\right)-\cos(\phi)\right)(x^{2}+x_{0}^2)\coth\left(2\sqrt{2}\pi r t\right)\right)\\
&&\frac{\exp\left(-4\pi^{2}r^{2}t\right)|(2\pi^{2}r^{2})^{\frac{1}{4}}|}{\sqrt{\left(4\pi t\right)^{2}2\pi\sinh\left(2\sqrt{2}\pi rt\right)}}\exp\left(-\frac{(x-x_{0})^{2}}{4t}\left(1+\sin(2\theta)\right)\right)\\
&&\exp\left(\frac{xx_{0}\left(\sin\left(\phi\right)\left(\cos\left(\theta\right)-\sin\left(\theta\right)\right)-\cos\left(\phi\right))\right)\sqrt{2}\pi r}{\sinh{\left(2\sqrt{2}\pi r t\right)}}\right)\\
&&\exp\left(-\frac{(x-x_{0})^{2}}{4t}\left(\frac{\left(\sin\left(\theta\right)-\cos\left(\theta\right)\right)\cos\left(\phi\right)-\sin\left(\phi\right)}{r\cos\left(\phi\right)}\right)^{2}\right)
\end{eqnarray*}

\subsection{Large $x$, Large $y$ asymptotics of the skewness for line source initial data}
Here, we derive the asymptotics for the skewness along the line $x=y$ for large values of $y$ and establish that the skewness is positive for all time in this limit.

\subsubsection{Mean,$\langle T\rangle$}
\begin{eqnarray*}
\langle T\rangle=\int_{{R}}dk\frac{\exp\left(2\pi i k y\right) }{\sqrt{\cosh\left(2\sqrt{2}\pi kt\right)}}\exp\left(-4\pi^{2}k^{2}t\right)\exp\left(-\frac{\pi}{\sqrt{2}}|k|x^{2}\tanh{\left(2\sqrt{2}\pi k t\right)}\right)
\end{eqnarray*}
After setting $x=y$ and $u=ky,$ then
\begin{eqnarray*}
&&\langle T\rangle=\frac{1}{y} \int_{R}{du\frac{\exp\left(2\pi i u\right)}{\sqrt{\cosh\left(\frac{2\sqrt{2}\pi u t}{y}\right)}}\exp\left(-\frac{4\pi^{2}u^{2}t}{y^{2}}\right)\exp\left(-\frac{\pi |u|y^{2}\tanh\left(\frac{2\sqrt{2}\pi |u| t}{y}\right)}{\sqrt{2}y}\right)}\\
&&\sim \frac{1}{y}\int_{R}\exp\left(2\pi i u\right)\exp\left(-2\pi^{2}u^{2}t\right)=\frac{1}{y\sqrt{2\pi t}}\exp\left(\frac{1}{2t}\right)\quad\text{as}\quad y\rightarrow\infty
\end{eqnarray*}
Note that we used $\tanh(\bar u)\sim \bar u$ for small $\bar u.$
\subsubsection{Second Field Moment, $\langle T^{2}\rangle$}
In this case we will make the following change of variables: 
\begin {eqnarray*}
\bar{u}=uy,u=k_{1}+k_{2},v=k_1-k_2
\end{eqnarray*}
Note that $u^{2}+v^2=2(k_1^2+k_2^2)$
Then \begin{eqnarray*}
k_1=\frac{u+v}{2},k_2=\frac{u-v}{2},|J|=\frac{\partial (k_1,k_2)}{\partial (u,v)}=\frac{-1}{2}
\end{eqnarray*}
\begin{eqnarray*}
&&\langle T^{2}\rangle= \int_{R^{2}}dk_{1}dk_{2}{\frac{\exp\left(2\pi i (k_{1}+k_{2})y\right)}{\sqrt{\cosh\left (2\sqrt{2}\pi( \sqrt{k_{1}^{2}+k_{2}^{2}})t\right)}}\exp\left(-4\pi^{2}(k_1^{2}+k_2^{2})t\right)}\\
&&{\exp\left(-\frac{\pi}{\sqrt{2}\sqrt{k_1^2+k_2^2}}(k_1+k_2)^{2}x^2\tanh\left(2\sqrt{2}\pi\sqrt{k_1^2+k_2^2}t\right)\right)}\\
&&=\frac{1}{2y}\int_{R^2}{d\bar {u}dv}\frac{\exp\left(2\pi i \bar{u}\right)}{\sqrt{\cosh{\left(2\pi t \sqrt{\frac{\bar{u}^{2}}{y^{2}}+v^{2}}\right)}}}\exp\left(-2\pi^{2}\left(\frac{\bar{u}^{2}}{y^{2}}+v^{2}\right)\right)\\
&&{\exp\left(-\frac{\pi}{\sqrt{\frac{\bar {u}^{2}}{y^{2}}+v^{2}}}\bar{u}^{2}\tanh{\left(2\pi \sqrt{\frac{\bar{u}^{2}}{y^{2}}+v^{2}}\right)}t\right)}\\
&&\sim \frac{1}{2y}\int_{R}dv\int_{R}d\bar{u}\frac{\exp\left(-2\pi^{2}v^{2}\right)}{\sqrt{\cosh\left(2\pi t |v|\right)}}\exp\left(2\pi i \bar{u}-\frac{\pi \tanh\left(2\pi |v|t\right)\bar{u}^{2}}{|v|}\right)\\
&&\sim\frac{1}{2y}\int_{R}{dv \frac{\sqrt{|v|}\exp\left({-2\pi^{2}}v^{2}-{\pi|v|\coth\left(2\pi |v|t\right)}\right)}{\sqrt{\sinh\left(2\pi |v|t\right)}}}\quad\text{as}\quad y\rightarrow\infty
\end{eqnarray*}
\subsubsection{Third Field Moment, $\langle T^{3}\rangle$}
If we make the following change of variables:
\begin{eqnarray*}
&&u=k_1+k_2+k_3\\
&&v=k_1-k_2\\
&&w=-k_1-k_2+2k_3
\end{eqnarray*}
and $\bar{u}=uy, x=y.$
Then we have
\begin{eqnarray*}
\frac{\partial(k_1,k_2,k_3)}{\partial (u,v,w)}=-\frac{1}{6}\quad\text{and}\quad k_1^{2}+k_2^2+k_3^2=\frac{u^2}{3}+\frac{v^2}{2}+\frac{w^2}{6}
\end{eqnarray*}
\begin{eqnarray*}
&&\langle T^{3}\rangle=\int_{R^3}{dk_1dk_2dk_3 }{\frac{\exp\left(2\pi i \left(k_1+k_2+k_3\right)y\right)\exp\left(-4\pi^{2}\left(k_1^2+k_2^2+k_3^2\right)t\right)}{\sqrt{\cosh\left(2\sqrt{2}\pi \sqrt{k_1^2+k_2^2+k_3^2}t\right)}}}\\
&&\exp\left(-\frac{\pi}{\sqrt{2\left(k_1^2+k_2^2+k_3^2\right)}}\left(k_1+k_2+k_3\right)^{2}x^{2}\tanh\left(2\sqrt{2}\pi\sqrt{k_{1}^2+k_{2}^{2}+k_{3}^{2}}t\right)\right)\\
&&=\frac{1}{6y}\int_{R^{3}}{d\bar{u}dvdw} {\frac{\exp\left(2\pi i {\bar{u}}\right)}{\sqrt{\cosh\left(2\sqrt{2}\pi t\sqrt{\frac{\bar{u}^{2}}{3y^{2}}+\frac{v^{2}}{2}+\frac{w^{2}}{6}}\right)}}}{\exp\left(-4\pi^{2}\left(\frac{\bar{u}^{2}}{3y^2}+\frac{v^{2}}{2}+\frac{w^{2}}{6}\right)t\right)}\\
&&\exp\left(-\frac{\pi \bar{u}^{2}}{\sqrt{2\left(\frac{\bar{u}^{2}}{3y^{2}}+\frac{v^{2}}{2}+\frac{w^{2}}{6}\right)}}\tanh\left(2\sqrt{2}\pi t\sqrt{\frac{\bar{u}^{2}}{3y^{2}}+\frac{v^{2}}{2}+\frac{w^{2}}{6}}\right)\right)\\
&&\sim \frac{1}{6y}\int_{R^3}d\bar{u}dvdw{\frac{\exp\left(2\pi i \bar{u}\right)}{\sqrt{\cosh\left(2\sqrt{2}\pi t\sqrt{\frac{v^{2}}{2}+\frac{w^{2}}{6}}\right)}}\exp\left(-4\pi^{2}\left(\frac{v^{2}}{2}+\frac{w^{2}}{6}\right)t\right)}\\
&&\exp\left(-\frac{\pi \bar{u}^{2}}{\sqrt{2\left(\frac{v^{2}}{2}+\frac{w^{2}}{6}\right)}}\tanh\left(2\sqrt{2}\pi t \sqrt{\frac{v^{2}}{2}+\frac{w^{2}}{6}}\right)\right)\\
&&\sim \frac{1}{6y}\int_{R^{2}}{dvdw}{\frac{\sqrt{\sqrt{2\left(\frac{v^{2}}{2}+\frac{w^{2}}{6}\right)}}}{\sqrt{\sinh\left(2\sqrt{2}\pi t\sqrt{\frac{v^{2}}{2}+\frac{w^{2}}{6}}\right)}}\exp\left(-4\pi^{2}\left(\frac{v^{2}}{2}+\frac{w^{2}}{6}\right)t\right)}\\
&&\exp\left( -\pi\sqrt{2\left(\frac{v^{2}}{2}+\frac{w^{2}}{6}\right)}\coth\left(2\sqrt{2}\pi t\sqrt{\frac{v^{2}}{2}+\frac{w^{2}}{6}}\right)\right)\quad \text{as}\quad y\rightarrow \infty
\end{eqnarray*}
Note that 
\begin{eqnarray*}
Var\left(T\right)&=&\langle T^{2}\rangle-\langle T\rangle^{2}\\
&\sim& \langle T^{2}\rangle \quad\text{as}\quad y\rightarrow \infty
\end{eqnarray*}
\begin{eqnarray*}
Sk(x,y,t)=\frac{\langle T^{3}\rangle-3\langle T^{2}\rangle \langle T\rangle+2\langle T\rangle }{\left (\langle T^{2}\rangle -\langle T\rangle ^{2}\right) ^{\frac{3}{2}}}\sim \frac{\langle T^{3}\rangle}{\left(\langle T^{2}\rangle\right)^{\frac{3}{2}}}>0\quad \text{as}\quad y\rightarrow \infty
\end{eqnarray*}
\subsection{Small Peclet (short time) asymptotics of the skewness}
Here, we derive the small Peclet (or equivalently, small time) asymptotic expansion of the skewness for both types of initial data along the $y$-axis.  

\subsubsection{Line Source Analysis}
This asymptotic expansion is directly obtained by Taylor expansion of the hyperbolic cosine function in the Mehler Kernels:
\begin{eqnarray*}
&&\frac{1}{\sqrt{\cosh\left(2\sqrt{2}\pi Pe |\mathbf{k}|t\right)}}\sim -\frac{556}{45} \pi ^6 |\mathbf{k}|^6 {Pe}^6 t^6+\frac{14}{3} \pi ^4 |\mathbf{k}|^4 {
Pe}^4 t^4-2 \pi ^2 |\mathbf{k}|^2 {Pe}^2 t^2+1\\
&&\text{as}\quad t\rightarrow 0^+
\end{eqnarray*}
Then we obtain through direct integration the first moment for line source initial data:
\begin{eqnarray*}
\langle \hat{T} \rangle^{L}&=e^{-4\pi^{2}k^{2}t}\frac{e^{-\frac{\pi{Pe}k{\tanh(2\sqrt{2}\pi k {Pe}t)}x^{2}}{\sqrt{2}}}}{\sqrt{\cosh(2\sqrt{2}\pi k {Pe}t)}}
\end{eqnarray*}
\begin{eqnarray*}
&&{\langle T(t,0,y)\rangle^{L}}\sim
-\frac{139 {Pe}^6 e^{-\frac{y^2}{4 t}} \left(120 t^3-180 t^2 y^2+30 t
   y^4-y^6\right)}{92160 \sqrt{\pi } \sqrt{t}}\\
   &&+\frac{7 {Pe}^4 e^{-\frac{y^2}{4 t}} \left(12 t^2-12 t y^2+y^4\right)}{768 \sqrt{\pi } \sqrt{t}}
   -\frac{{Pe}^2 e^{-\frac{y^2}{4 t}} \left(2 t-y^2\right)}{16 \sqrt{\pi }
   \sqrt{t}}+\frac{e^{-\frac{y^2}{4 t}}}{2 \sqrt{\pi } \sqrt{t}}\qquad \\
   &\text{as}\quad t\rightarrow 0^{+}
\end{eqnarray*}
Next, we compute the second Moment with line source initial data:
\begin{eqnarray*}
\langle \hat{T}^{2}(\mathbf{x},\mathbf{k},t) \rangle^{L} =&\frac{e^{-\frac{\pi{Pe}|\mathbf{k}|}{\sqrt{2}}\big(\frac{x_{1}k_{1}+x_{2}k_{2}}{|\mathbf{k}|}\big)^{2}\tanh(2\sqrt{2}\pi |\mathbf{k}| {Pe}t)}}{\sqrt{\cosh(2\sqrt{2}\pi |\mathbf{k}| {Pe}t)}}e^{-4\pi^{2}|\mathbf{k}|^{2}t}
\end{eqnarray*}
\begin{eqnarray*}
&&{\langle T^{2}(t,0,y)\rangle^{L}}\sim
-\frac{139 {Pe}^6 e^{-\frac{y^2}{2 t}} \left(48 t^3-72 t^2 y^2+18 t y^4-y^6\right)}{23040 \pi  t}\\
&&+\frac{7
   {Pe}^4 e^{-\frac{y^2}{2 t}} \left(8 t^2-8 t y^2+y^4\right)}{384 \pi  t}
   -\frac{{Pe}^2 e^{-\frac{y^2}{2 t}} \left(2 t-y^2\right)}{16 \pi  t}+\frac{e^{-\frac{y^2}{2 t}}}{4 \pi  t}\\
   &&\qquad \text{as}\quad t\rightarrow 0^{+}
\end{eqnarray*}
\begin{eqnarray*}
&&Var(T)=\langle T^{2}\rangle^{L}-\big(\langle T\rangle^{L}\big)^{2}\\
&&\sim\frac{{Pe}^6 e^{-\frac{y^2}{2 t}} \left(-209664 t^3+314496 t^2 y^2-122304 t y^4+8736 y^6\right)}{2580480 \pi  t}+\frac{{Pe}^4 e^{-\frac{y^2}{2 t}} \left(y^{2}-2t\right)^{2}}{192 \pi  t}\\
&& \text{as}\quad t\rightarrow 0^{+}
\end{eqnarray*}
We should note that the asymptotics converge to $0$ as $t\rightarrow 0.$, and this function is a non-negative quantity. \\

Lastly, we compute the short time asymptotics for the third moment with line source initial data:
\begin{eqnarray*}
\langle \hat{T}^{3}(\vec{x},\mathbf{k},t) \rangle^{L} =&\frac{e^{-\frac{\pi{Pe}|\mathbf{k}|}{\sqrt{2}}(z_{1}^{2})\tanh(2\sqrt{2}\pi |\mathbf{k}| {Pe}t)}}{\sqrt{\cosh(2\sqrt{2}\pi |\mathbf{k}| {Pe}t)}}e^{-4\pi^{2}|\mathbf{k}|^{2}t}
\end{eqnarray*}

\begin{eqnarray*}
&&{\langle T^{3}(t,0,y)\rangle^{L}}\sim-\frac{139 {Pe}^6 e^{-\frac{3 y^2}{4 t}} \left(280 t^3-420 t^2 y^2+126 t y^4-9
   y^6\right)}{122880 \pi ^{3/2} t^{3/2}}\\
   &&+\frac{7 {Pe}^4 e^{-\frac{3 y^2}{4 t}} \left(20 t^2-20 t y^2+3 y^4\right)}{1024 \pi ^{3/2} t^{3/2}}-\frac{3 {Pe}^2 e^{-\frac{3 y^2}{4 t}} \left(2 t-y^2\right)}{64 \pi ^{3/2}
   t^{3/2}}
   +\frac{e^{-\frac{3 y^2}{4 t}}}{8 \pi ^{3/2} t^{3/2}}\\
   && \text{as}\quad t\rightarrow 0^{+}
\end{eqnarray*}

\begin{eqnarray*}
&\langle{T^{3}}\rangle^{L}-3\langle T^{2}\rangle^{L}\langle T\rangle^{L}+2\big(\langle T\rangle^{L}\big)^{3}\sim -\frac{{Pe}^6 e^{-\frac{3 y^2}{4 t}} \left(2 t-y^2\right)^3}{960 \pi ^{3/2} t^{3/2}}\qquad \text{as}\quad t\rightarrow 0^{+}
\end{eqnarray*}
Constructing the short time skewness, we obtain: 
\begin{eqnarray*}
Sk^{L}\sim  \frac{8\sqrt{3}}{5}\sgn{(y^2-2 t)}\quad\text{as}\quad t\rightarrow 0^{+}\;.
\end{eqnarray*}

\subsubsection{Point Source Analysis}
We repeat the prior calculation for the case of the point source initial conditions.

We find the first moment for point source initial data:
\begin{eqnarray*}
\langle \hat{T} \rangle^{P}=e^{-4\pi^{2}k^{2}t}(2\pi^{2}k^{2}{Pe}^{2})^{\frac{1}{4}}\frac{e^{-\frac{\pi{Pe}k{\coth(2\sqrt{2}\pi k {Pe}t)}(x^{2}+x_{0}^{2})}{\sqrt{2}}}}{\sqrt{2\pi\sinh(2\sqrt{2}\pi k {Pe}t)}}e^{\frac{\sqrt{2}{(\pi k{Pe})}xx_{0}}{\sinh(2\sqrt{2}\pi k {Pe}t)}}
\end{eqnarray*}
\begin{eqnarray*}
&&{\langle T(t,0,y)\rangle^{P}}\sim
\frac{61 {Pe}^6 e^{-\frac{y^2}{4 t}} \left(120 t^3-180 t^2 y^2+30 t
   y^4-y^6\right)}{3870720 \pi  t}\\
   &&+\frac{{Pe}^4 e^{-\frac{y^2}{4 t}} \left(12
   t^2-12 t y^2+y^4\right)}{2560 \pi  t}+\frac{{Pe}^2 e^{-\frac{y^2}{4 t}} \left(2
   t-y^2\right)}{96 \pi  t}+\frac{e^{-\frac{y^2}{4 t}}}{4 \pi  t}\\
   &\qquad \text{as}\quad t\rightarrow 0^{+}
\end{eqnarray*}
Next, we compute the short time asymptotics for the second moment with point source initial data where $z_{1}=\frac{x_{1}k_{1}+x_{2}k_{2}}{|\mathbf{k}|},$ $z_{2}=\frac{-x_{1}k_{2}+x_{2}k_{1}}{|\mathbf{k}|}$:
\begin{eqnarray*}
\langle \hat{T}^{2}(\vec{x},\mathbf{k},t) \rangle^{P} =&\frac{e^{-\frac{\pi{Pe}|\mathbf{k}|}{\sqrt{2}}\big(z_{1}^{2}+z_{0}^{2}\big)\coth(2\sqrt{2}\pi |\mathbf{k}| {Pe}t)}e^{-\frac{\sqrt{2}\pi|\mathbf{k}|{Pe}(z_{1}z_{0})}{\sinh(2\sqrt{2}\pi |\mathbf{k}| {Pe}t)}}}{\sqrt{2\pi \sinh(2\sqrt{2}\pi |\mathbf{k}| {Pe}t)}}e^{-4\pi^{2}|\mathbf{k}|^{2}t}\frac{(2\pi^{2}|\mathbf{k}|^{2} {Pe}^{2})^{\frac{1}{4}}e^{-\frac{(z_{2}-z_{0})^{2}}{4t}}}{\sqrt{4\pi t}}
\end{eqnarray*}
\begin{eqnarray*}
&&\langle T(t,0,y)^{2}\rangle^{P}\sim-\frac{61 {Pe}^6 e^{-\frac{y^2}{2 t}} \left(48 t^3-72 t^2 y^2+18 t y^4-y^6\right)}{1935360 \pi ^2 t^2}\\
&&+\frac{{Pe}^4
   e^{-\frac{y^2}{2 t}} \left(8 t^2-8 t y^2+y^4\right)}{2560 \pi ^2 t^2}-\frac{{Pe}^2 e^{-\frac{y^2}{2 t}} \left(2
   t-y^2\right)}{192 \pi ^2 t^2} +\frac{e^{-\frac{y^2}{2 t}}}{16 \pi ^2 t^2}\\
&&\text{as}\quad t\rightarrow 0^{+}
\end{eqnarray*}
\begin{eqnarray*}
Var(T)=\langle T^{2}\rangle^{P}-\big(\langle T\rangle^{P}\big)^{2}\sim\frac{{Pe}^4 e^{-\frac{y^2}{2 t}} \left(2 t-y^2\right)^2}{11520 \pi ^2 t^2}-\frac{{Pe}^6 e^{-\frac{y^2}{2 t}} \left(24
   t^3-36 t^2 y^2+14 t y^4-y^6\right)}{64512 \pi ^2 t^2}
\end{eqnarray*}
 $ \text{as}\quad t\rightarrow 0^{+}.$\\

Lastly, we find the short time asymptotics for the third moment for point source initial data where $z_{1}=\frac{x_{1}k_{1}+x_{2}k_{2}+x_{3}k_{3}}{|\mathbf{k}|},$ $z_{2}=\frac{-x_{1}k_{2}+x_{2}k_{1}}{\sqrt{k_{1}^{2}+k_{2}^{2}}},$ $z_{3}=\frac{-k_{1}k_{3}x_{1}-k_{2}k_{3}x_{2}}{|\mathbf{k}|\sqrt{k_{1}^{2}+k_{2}^{2}}}+\frac{x_{3}\sqrt{k_{1}^{2}+k_{2}^{2}}}{|\mathbf{k}|}$:
\begin{eqnarray*}
\langle \hat{T}^{3}(\vec{x},\mathbf{k},t) \rangle^{P} =&\frac{e^{-\frac{\pi{Pe}|\mathbf{k}|}{\sqrt{2}}\big(z_{1}^{2}+z_{0}^{2}\big)\coth(2\sqrt{2}\pi |\mathbf{k}| {Pe}t)}e^{-\frac{\sqrt{2}\pi|\mathbf{k}|{Pe}(z_{1}z_{0})}{\sinh(2\sqrt{2}\pi |\mathbf{k}| {Pe}t)}}}{\sqrt{2\pi \sinh(2\sqrt{2}\pi |\mathbf{k}| {Pe}t)}}e^{-4\pi^{2}|\mathbf{k}|^{2}t}\frac{(2\pi^{2}|\mathbf{k}|^{2} {Pe}^{2})^{\frac{1}{4}}e^{-\frac{(z_{2}-z_{0})^{2}}{4t}}e^{-\frac{(z_{3}-z_{0})^{2}}{4t}}}{(\sqrt{4\pi t})^{2}}
\end{eqnarray*}
\begin{eqnarray*}
&&\langle T^{3}(t,0,y)\rangle^{P}\sim-\frac{61 {Pe}^6 e^{-\frac{3 y^2}{4 t}} \left(280 t^3-420 t^2 y^2+126 t y^4-9 y^6\right)}{20643840 \pi ^3 t^3}\\
&&+\frac{3
   {Pe}^4 e^{-\frac{3 y^2}{4 t}} \left(20 t^2-20 t y^2+3 y^4\right)}{40960 \pi ^3 t^3}
  -\frac{{Pe}^2 e^{-\frac{3 y^2}{4
   t}} \left(2 t-y^2\right)}{512 \pi ^3 t^3}+\frac{e^{-\frac{3 y^2}{4 t}}}{64 \pi ^3 t^3}\\\
   &&\text{as}\quad t\rightarrow 0^{+}
\end{eqnarray*}
\begin{eqnarray*}
&\langle{T^{3}}\rangle^{P}-3\langle T^{2}\rangle^{P}\langle T\rangle^{P}+2\big(\langle T\rangle^{P}\big)^{3}\sim -\frac{{Pe}^6 e^{-\frac{3 y^2}{4 t}} \left(2 t-y^2\right)^3}{483840 \pi ^3 t^3}\qquad \text{as}\quad t\rightarrow 0^{+}
\end{eqnarray*}
For the case of $x=y=x_{0}=0,$
\begin{eqnarray*}
Sk^{P}\sim  \frac{8\sqrt{5}}{7}\sgn{(y^2-2 t)}\quad\text{as}\quad t\rightarrow 0^{+}\;.
\end{eqnarray*}
 \subsection{Evaluation of Numerical Integrals}
 The single point statistics at $(x=0,y=0)$ for the both point and line source initial data for \ref{passivescalar} rely on the exact calculation of the pdf of the functionals of $L2-$ norm of brownian motion and the brownian bridge. When we do this analysis, we used the built in numerical integrator in Mathematica. However, our numerical experiments led us to take $\tt AccuracyGoal=4,$ $\tt MinRecursion=1,$ $\tt MaxRecursion=20,$ $\tt Method=LevinRule.$
\end{footnotesize}


\begin{thebibliography}{99}
\bibitem{aminiancamassamcl} 
Aminian, M.,  Camassa,R., McLaughlin, R.M., ``{ Mass distribution and skewness for passive scalar transport in pipes with polygonal and smooth cross-sections }", {\em arxiv}, {\bf } (2017) . 
\bibitem{PRL}
 Aminian,M. , Bernardi, F. ,Camassa,R., McLaughlin, R. M. , ``Squaring the Circle: Geometric Skewness and Symmetry Breaking for Passive Scalar Transport in Ducts and Pipes", {\em Phys. Rev. Lett.} {\bf 115}, 154503 (2015).

\bibitem{Science}
 Aminian,M. , Bernardi, F. ,Camassa, R., Harris, D. M., McLaughlin, R. M. , ``How boundaries shape chemical delivery in microfluidics", \textit{Science},  {\bf 
354 (6317)}, 1252-1256.  

\bibitem{Anderson}
Anderson, T.W. - Darling, D.A.: 
``Asymptotic Theory of certain "goodness of fit" criteria based on stochastic processes.", \textit{Ann. Math. Statist.}, 1952, \textbf{23}, pp. 193-212.

\bibitem{Barton}  Barton, N.G.:, ``On the method of moments for solute dispersion", {\em J. Fluid Mech.}, {\bf 126} 205-218, 1983.  

\bibitem{CamassaLinMcL1}
Bronski, J.C. ,  Camassa,R.,  Lin, Z., McLaughlin, R.M., and Scotti, A.
``An explicit family of probability measures for passive scalar diffusion in a random flow", {\em J. Stat. Phys.}, {\bf 128} (2007), 927--968. 

\bibitem{BronskiMcL1}
Bronski, J.C. and  McLaughlin, R.M., `` Passive scalar intermittency and the ground state of Schrodinger operators", {\em Phys. Fluids},  {\bf 9} (1997), 181--190. 


\bibitem{BronskiMcL2}
Bronski, J.C. and  McLaughlin, R.M.,``Rigorous estimates of the tails of the probability distribution function for the random linear shear model", {\em J. Stat. Phys.},  {\bf 98} (2000), 897--915. 


\bibitem{CamassaLinMcL2}
 Camassa,R.,  Lin, Z., McLaughlin, R. M.,
``Evolution of the probability measure for the Majda model: New invariant measures and breathing PDFs",
{\em J. Stat. Phys. },  {\bf 130} (2008) 343--371.



\bibitem{CamassaLinMcLMajda} 
 Camassa,R.,  Lin, Z., McLaughlin, R. M.,``The exact evolution of the scalar variance in 
pipe and channel flow", {\em Commun. Math. Sci.}, {\bf 8} (2010) 601--626. 


\bibitem{kac}
Kac, M.: ``On the average of a certain Wiener functional an related limit theorem in calculus of probability", \textit{Trans. Amer. Math. Soc.}, 1946, \textbf{3}, pp. 762-772.
\bibitem{Kilic}
Kilic, Z.: ``Random transport of a passive scalar", \textit{Ph. D. Thesis}, Applied Mathematics Program,UNC at Chapel Hill, 2018.

\bibitem{Kraichnan}
R.H. Kraichnan, ``Small-scale structure of a scalar field convected by turbulence",  {\em Phys. Fluids}, {\bf 11} (1968), 945-953 

\bibitem{Libchaber}
Castaing, B. , Gunaratne, G. , Heslot,F.,  Kadanoff, L. ,  Libchaber, A. ,  Thomae, S.,  Wu, X-Z. , Zaleski, S. , and  Zanetti, G., 
\newblock ``Scaling of hard thermal turbulence in Rayleigh-Benard convection", 
\newblock {\it J. Fluid Mech.}, {\bf 204} (1989), 1--30. 

\bibitem{Majda}
Majda, A.J.: ``The random uniform shear layer : an explicit example of turbulent diffusion with broad tail probability distributions" \textit{Phys. Fluids A}, 1993, \textbf{5}, pp. 1963-1970.

\bibitem{4}
Majda, A.J.: ``Explicit inertial range renormalization theory in a model for turbulent diffusion", \textit{J. Stat. Phys.}, 1993, \textbf{73}\textbf{3-4}, pp. 515-542.

\bibitem{5}
Majda, A.J, Kramer, P.: ``Simplified models for turbulent diffusion: Theory, numerical modelling, and physical phenomenon", \textit{Phys. Reports}, 1999, \textbf{314}, pp. 237-574.

\bibitem{marsaglia64}
 Marsaglia,G.,  and  Bray, T. A., T..: `` A convenient method for generating normal variables",\textit{SIAM Rev}, 1964, \textbf{5}, pp. 260-264.
 
 \bibitem{matsumoto98}
Matsumoto,M.,  and  Nishimura, T.: `` Mersenne twister: a 623-dimensionally equidistributed uniform pseudo-random number generator", \textit{TOMACS}, 1998, \textbf{8}, pp. 3-30.

\bibitem{McLMajda}
McLaughlin, R. M., Majda, A.J.: ``An explicit example with non-Gaussian probability distribution for nontrivial scalar mean and fluctuation", \textit{Phys. Fluids}, 1996, \textbf{8}, pp. 536-547.

\bibitem{McLthesis}
McLaughlin, R. M.: ``Turbulent diffusion", \textit{Ph. D. thesis}, Program in Applied and Computational Mathematics,Princeton University,1994.

\bibitem{Pope}
Pope, S.B.: ``Mapping Closures for Turbulent Mixing and Reaction", \textit{Theoret. Comp. Fluid Dyanmics}, 1991, \textbf{2}, pp. 255-270.


\bibitem{Smirnov}
Smirnov, N.V.: ``On the distribution of the von Mises $\omega^{2}-$criterion (in Russian). Matem Sbornik", \textit{Annals of Prob.}, 1937, \textbf{5}, pp. 973-993.

\bibitem{Vanden}
Vanden-Eijnden, E.: ``Non-Gaussian invariant measures for the Majda model of decaying turbulent transport", \textit{Comm. Pure Appl. Math.}, \textbf{54(9)},2001, pp. 1146-1167.

\end{thebibliography}
\end{document}